\documentclass[12pt]{iopart} 
\usepackage{epsfig} 
\def\bit{\begin{itemize}} 
\def\fit{\end{itemize}} 
\def\beq{\begin{equation}} 
\def\feq{\end{equation}} 
\def\beqar{\begin{eqnarray}}
\def\feqar{\end{eqnarray}}
\def\benu{\begin{enumerate}} 
\def\fenu{\end{enumerate}} 

\begin{document} 
\sloppy   
\newpage  

\title{Recent advances in neutrinoless double beta decay search}  

\author{Lino Miramonti\footnote[1]{e-mail:miramonti@mi.infn.it}} 
\address{Physics Department of Milano University\\ \&\\ 
National Institute of Nuclear Physics (INFN) of Milano\\
Via Celoria 16, 20133 Milano (Italy)}  

\author{Franco Reseghetti\footnote[2]{e-mail: reseghetti@santateresa.enea.it}  
                           } 
\address{Italian National Agency for New Technologies, Energy and the Environment (ENEA) \\
Loc. S.Teresa, 19036 Pozzuolo di Lerici (Italy)}  

\begin{abstract}
Even after the discovery of neutrino flavour oscillations, based on data from atmospheric, solar, reactor, 
and accelerator experiments, many characteristics of the neutrino remain unknown. 
Only the neutrino square-mass differences and the mixing angle values have been estimated, while the value of each
mass eigenstate still hasn't. Its nature (massive Majorana or Dirac particle) is still escaping. \\
Neutrinoless double beta decay ($0\nu$-DBD) experimental discovery could be
the ultimate answer to some delicate questions of elementary particle and nuclear physics. 
The Majorana description of neutrinos allows the $0\nu$-DBD process, and consequently either a mass value 
could be measured or the existence of physics beyond the standard should be  
confirmed without any doubt. As expected, the $0\nu$-DBD measurement is a very difficult field of application for 
experimentalists. \\
In this paper, after a short summary of the latest results in neutrino physics, 
the experimental status, the R\&D projects, and perspectives in $0\nu$-DBD sector are reviewed.  
\end{abstract} 

\tableofcontents  
  
\maketitle 

\section{Introduction}
Recent experimental results and analyses from atmospheric, solar, accelerator and reactor 
neutrino physics, 
\cite{Ahmad2004, Ashie2004, Ishitsuka2004, Araki2004, Nakaya2004, Nakahata2004}, indicate
that neutrinos change their flavour, and, consequently, do have a mass. Unfortunately, in the 
oscillation experiments only the square-mass differences between pairs of flavours and the mixing angle can be 
estimated, while the magnitude of the masses still remains unknown. 

A fundamental question arises: Is the neutrino coincident with its own anti-particle? 
If the answer is positive, a neutrino is a Majorana massive particle, if not, a 
Dirac massive fermion.

This is a crucial point. If neutrinos and anti-neutrinos are identical, the balance 
between particles and anti-particles in Early Universe could have been modified, leading to 
the asymmetry between matter and anti-matter. 
Consequently, the Majorana nature of neutrinos is linked to the observed baryon asymmetry. 

A possible answer is given by the $0\nu$-DBD process, which violates the lepton number 
by two units, and it is forbidden if neutrino differs from anti-neutrino. Such a reaction is a unique tool to measure 
the Majorana neutrino phases, and to determine the absolute neutrino mass scale.
The double beta decay (DBD) is a very rare nuclear transition firstly described in the 30's by 
Maria Goeppert-Mayer, \cite{Mayer1935}, who estimated the half-life of the process to be very long. 
Few years later, Majorana proposed a fermion two-component theory alternative to the Dirac description, 
\cite{Majorana1937}. Racah also described the possibility of the transformation of two neutrons into two protons, 
with the emission of 2 electrons, but without neutrinos, \cite{Racah1937}. Then, Furry quoted shorter half-life for 
$0\nu$-DBD reactions, \cite{Furry1939}. 

We remind that such two-component fermions are called Majorana particles, whereas Dirac fermions are four-component 
particles. 

DBD is a second order weak semileptonic spontaneous nuclear transition in which nuclear electric charge changes by 
two units, whereas the mass number remains unchanged. Sometimes, two nucleons (protons or neutrons) simultaneously 
emit a lepton pair each, that has been observed in a number of experiments. Such a  process occurs only if the 
parent decaying nucleus is less bound that the final one, the intermediate nucleus being less bound than both these nuclei.  

Different cases are allowed, with or without the emission of neutrinos, and of other particles. 
Many nuclei undergo these processes: the paring force makes the even-even nuclei, which have an even number both 
of protons and neutrons, more stable than odd-odd nuclei, with broken pairs. The usual beta decay transition from an 
even-even parent nucleus to a neighbouring odd-odd nucleus ((A,Z)$\rightarrow$(A, Z+1)) is energetically forbidden, 
whereas the DBD transition to the daughter nucleus (A, Z+2) is allowed. 

DBD process naturally occurs for a few tenths of even-even nuclei, mainly from initial ground to final ground
states, see \cite{Moe and Vogel 94,Tretyak and Zdesenko 95,Tretyak and Zdesenko 02}. In table 
\ref{compilation of dbd candidates}, a list of possible interesting reactions having a high Q value is shown. 
In the case of $^{48}$Ca and $^{96}$Zr, the standard beta decay is energetically allowed, but 
strongly suppressed because of the large difference in angular momentum $0^{+} \rightarrow 6^{+}$.

\begin{table*}
\centering
\caption[]{Compilation, ordered by following the atomic mass number, of DBD candidate nuclei 
with Q$_{\beta \beta}>$ 1.7 MeV. In last column, r shows the isotopic abundance fraction in percent.}
\label{compilation of dbd candidates}
\begin{tabular} {ccc}   \hline
 Isotope                & Q$_{\beta \beta }$ (keV) &  r (\%)\\
\hline
$^{48}$Ca              & 4272                    & 0.187              \\
$^{76}$Ge              & 2039                    & 7.61               \\
$^{82}$Se              & 2995                     & 8.73               \\
$^{96}$Zr              & 3350             & 2.80               \\ 
$^{100}$Mo             & 3034                   & 9.63               \\
$^{110}$Pd             & 2000            & 11.72              \\
$^{116}$Cd             & 2805            & 7.49               \\
$^{124}$Sn             & 2287            & 5.79               \\
$^{130}$Te             & 2529            & 34.08              \\
$^{136}$Xe             & 2468            & 8.87               \\
$^{148}$Nd             & 1929            & 5.7                \\
$^{150}$Nd             & 3367            & 5.6                \\
$^{160}$Gd             & 1730            & 22.86              \\
\hline
\end{tabular}
\end{table*}

It has to be pointed out that the second-order process is allowed if Q is positive, 
but, if a single beta decay parent nucleus is unstable, it is practically impossible to distinguish 
the DBD from the usual, and most intensive, beta decay, i.e. the background.

We underline that in the Standard Model of Electroweak Interaction and Particles, a massless Dirac 
neutrino is introduced, and the $2\nu$-DBD reaction is allowed. Such a theory was very successful in all tested 
applications, and represented 
the most economical theory explaining weak and electromagnetic interactions, but experimental results, 
mainly from the neutrino sector, have shown its inadequacy. Moreover, it gives no answer 
to a lot of fundamental questions. Among them, we mention the "anomalous" difference between 
neutrino and the corresponding charged lepton mass, and the neutrino nature (Dirac or a Majorana particle).

The deficit of the solar neutrino flux and its complicated energy dependence is evident in all the known experiments. 
Atmospheric neutrinos show an anomalous ratio and an unexpected angular distribution between electronic and muonic 
components. Finally, the disappearance of neutrinos in a Japanese accelerator long baseline experiment, and the deficit in 
the measured anti-neutrino flux emitted by Japanese reactors, definitively confirm that neutrino flavour 
oscillations do occur, and, consequently, neutrinos do have a non-zero mass.

Therefore, DBD physics is still a fundamental field of application for both particle and nuclear physicists, 
as pointed out manu years ago in \cite{Haxton and Stephenson 84}. Its experimental detection  requires good technical advances, mainly looking for 
the background suppression. It also offers several constraints to the calculations of nuclear properties.
Taking into account the negligible decay rates, and problems with the evaluation of nuclear matrix elements, measured 
lifetimes agrees satisfactorily with theoretical values, and the process appears to be sufficiently understood.

We remember also that many good reviews, and dedicated papers on DBD processes have been prepared, 
see for instance \cite{Barabash2004, Elliott2004} and reference therein. 

We also mention the $Neutrinoless\, Double\, Beta\, Decay$ section of the NEUTRINO UNBOUND web page <http://www.nu.infn.it/> 
for a complete list of papers and links concerning the $0\nu$-DBD physics.

\section{Neutrino and Double Beta Decay}

\subsection{Neutrino mass and flavour oscillations}

Massive fermions are described by the four-component Dirac equation, in which left and right chirality eigenstates are
coupled. However, in the Standard Model of particles and interactions only left-handed neutrinos have interactions.

The neutrino Lagrangian has a Lorentz invariant mass term which includes three components. The first one 
(Dirac mass term) conserves the lepton quantum number and requires two chirality eigenstates: left and right.
The remaining ones (Majorana mass terms) violate the lepton number conservation, and each exists 
independently of the other element. Usually, two non-degenerate mass eigenvalues for each flavour are the result of the 
diagonalization of the more general Lagrangian. 

The most relevant case occurs when only the left mass N $\times$ N matrix is different from 0, where N is the number of 
neutrino flavours. Then, the unitary U matrix contains N$^2$ real parameters, N(N-1)/2 angles and N(N+1)/2 phases,  
whereas N terms represent unphysical phases. In reactions where the flavour lepton number changes, but not the total 
lepton number, such as in oscillation experiments, all mixing angles and (N-1)(N-2)/2 phases, which describe CP violation 
and the related possible oscillation probability differences between neutrinos and anti-neutrinos, can be computed. 
(N-1) phases can be calculated from processes like $0\nu$-DBD reactions, in which the total lepton number changes, 
but they have significance only for Majorana neutrinos. Practically, three CP violating phases appear in $U_{li}$ 
for Majorana particles \cite{Bilenky1980}. 
   
We assume that the flavour fields $\nu_{lL}$ ($l$ = e, $\mu$, $\tau$) are mixtures of the fields of three active neutrinos 
with definite masses, without the contribution of sterile neutrinos.
\begin{equation}
\nu_{lL} = \sum_{i=1}^{3}\,~ U_{li} \nu_{iL}, 
\label{002}
\end{equation}
where $\nu_{i}$ is the field of neutrino (Dirac or Majorana), and $U_{l i}$ is the Pontecorvo-Maki-Nakagawa-Sakata 
(PMNS) 3$\times$3 unitary left-handed lepton mixing matrix, which correlates the physical weak eigenstates  
($l=e, \mu, \tau$) to the mass eigenstates $m_i$ ($i$ = 1,2,3). 

The $U_{l i}$ matrix takes the following form in the standard representation:
\begin{equation}
U_{l i}= \left(
\begin{array}{ccc}
U_{e1}     & U_{e2}     & U_{e3}     \\
U_{\mu1}   & U_{\mu2}   & U_{\mu3}   \\
U_{\tau1}  & U_{\tau2}  & U_{\tau3}  \\
\end{array}
\right) 
\label{PMNS matrix}
\end{equation}
Therefore, the value of $\langle m_{\nu} \rangle$ depends on the value of the individual neutrino mass eigenstates 
$m_{i}$, the mixing matrix elements $U_{ei}$ of the first row, and the Majorana phases 
$\alpha_{i}$ (where $\alpha_{ij} = \alpha_{i}-\alpha_{j}$).
If three light massive Majorana neutrinos exist, the weak eigenstates $\nu_e$, $\nu_\mu$, and $\nu_\tau$ are a 
superposition of the mass eigenstates and the effective neutrino mass is:
\begin{equation}
\langle m_{\nu} \rangle^2 = \left| \sum_{i=1}^3 U^2_{ei}\,m_i \right|^2 = 
\left| \sum_{i=1}^3 |U_{ei}|^2 e^{i\alpha_{i1}} m_i \right|^2 
\label{neutrino mass}
\end{equation}
where two CP violating Majorana phases $\alpha_i$, which can be cancelled in the sum operation, are also included, and
the neutrino mass element $\left|\langle m^{\nu}_{ee} \rangle\right|$ is 
\begin{equation}
\left|\langle m^{\nu}_{ee} \rangle\right| = \left|\, 
m_1\left| U^2_{e1}\right|e^{i\alpha_1}\,+\,
m_2\left| U^2_{e2}\right|e^{i\alpha_2}\,+\,
m_3\left| U^2_{e3}\right|\,
\right|
\label{neutrinomasselement}
\end{equation}

Left-handed V-A weak currents and Majorana massive neutrinos are needed to describe the interactions occurring
in oscillation experiments. Newest results provide data which strongly constrains the mixing matrix elements 
and the differences in the square of masses eigenvalues, $\Delta m_{ij}^2 \equiv m_j^2 - m_i^2$. Terms with label \{1,2\}
describe solar mixing sector, whereas label \{2,3\} is concerning atmospheric mixing.

By now, we have clear evidence about neutrino oscillation and flavour mixing from the study of neutrino produced 
from different sources \cite{Giunti2003}.

\subsection{Experimental results}

At the begin of summer 2004, several analyses and new measurements have been presented by different
collaborations. The available results are shortly resumed.

\subsubsection{Atmospheric neutrino experiments}

Latest analyses on SuperKamiokande atmospheric neutrino dataset, \cite{Ashie2004,Ishitsuka2004}, which are based on 1489
live-days exposure and on enlarged fiducial volume for fully contained events (from 22500 to 26400 ton), confirm the deficit in
the muon neutrino flux with a strong zenithal dependence. The best explanation of such measurements is given in terms of a flavour
transition $\nu_{\mu} \rightarrow \nu_{\tau}$, with a squared mass difference 
$\Delta{m}^2_{\mathrm{atm}}$ in the range 
\begin{equation}
1.9 \cdot 10^{-3} < \Delta{m}^2_{\mathrm{atm}} < 3.0 \cdot 10^{-3} (eV)^2\ 
\label{range deltaM atm}
\end{equation}
at 90\% C.L., and a best fit value $\Delta{m}^2_{\mathrm {atm}}\simeq 2.4 \cdot 10^{-3} (eV)^2$.
The associated atmospheric mixing angle $\sin^2\,\vartheta_{\mathrm{atm}}$ is $\simeq$ 1.0 (i.e. maximal mixing angle), 
with the 90\% C.L. lower bound
\begin{equation}
\sin^2 2\,\vartheta_{\mathrm{atm}} > 0.90 \
 \label{sen2 teta atm}
\end{equation}

\subsubsection{Solar neutrino experiments}

The SuperKamiokande collaboration has recently presented an improved analysis on measurements before the incident (SK-I), and
preliminary results with the new detector setup (SK-II), at a higher energy threshold, \cite{Nakahata2004}. The 
obtained flux, which is $\sim$ 45\% of the predicted one, strenghtens the previous values. \\
The whole interaction rate for Gallium experiments, 68.10 $\pm$ 3.75 SNU,   
\cite{Cattadori2004}, is practically constant over a decade, even if data since 1997 could indicate a slightly reduced 
solar neutrino flux in latest years. 
SNO results indicate a great suppression of the flux for charged current and elastic scattering events, and
an indisturbed flux for neutral currents interactions, \cite{Ahmad2004}. \\
Consequently, the conversion from $\nu_{e}$ to 
$\nu_{\mu}$/$\nu_{\tau}$ arises as a natural explanation of the deficit and of the energy spectra, with 
\begin{equation}
\Delta{m}^2_{\mathrm{Sun}} \sim 7.0 \cdot 10^{-5} (eV)^2\ \qquad
\tan^2 \vartheta_{\mathrm{Sun}} \sim 0.42 \
\label{solar}
\end{equation} 

\subsubsection{Reactor neutrino experiments}

In June 2004, KamLAND collaboration has published the analysis of data collected between March 2002 and January 2004, 
with a slightly enlarged fiducial volume and improved detection setup, \cite{Araki2004, Gratta2004}. 
The expected anti-neutrino flux in absence of oscillation phenomena is 365.2 $\pm$ 23.7 events, 
whereas 258 events have been detected. This result is well explained if a flavour oscillation occurs; the best fit is obtained with
\begin{equation}
\Delta{m}^2_{12} = 8.3 \cdot 10^{-5}\, (eV)^2 \qquad \tan^2 \vartheta_{12} = 0.41\ 
\label{Kamland}
\end{equation}
The decay solution and the decoherence description are excluded at 95\% and 94\%, respectively.

\subsubsection{Long baseline accelerator neutrino experiments}

New results have been released by K2K collaboration, \cite{Nakaya2004}, after the installation of SCIBAR apparatus:
108 events have been detected, whereas the expected number of interactions is 150.9. As in previous dataset, the deficit 
is mainly due to $\mu$-like events (57 detected interactions instead 84.8 expected events). 
The best-fit solution in physical region is obtained with the following values:
\begin{equation}
\Delta{m}^2_{23} = 2.73 \cdot 10^{-3}\, (eV)^2 \qquad \sin^2 2\vartheta_{23} = 1.00\ 
\label{K2K}
\end{equation}
The neutrino oscillation is confirmed at 3.9 $\sigma$, whereas the significance of $\nu_\mu$ disappearance 
is at a level of 2.9 $\sigma$, and the energy spectrum distortion is at a level of 2.5 $\sigma$. 

Before latest experimental results from KamLAND, the mass values around $\sim 1.5 \cdot 10^{-4} (eV)^2$ 
(the high Large Mixing Angle solution) was not completely discarded, see for instance 
\cite{Bahcall2003, DeHolanda2003, Smy2003}. At the present moment, this mass solution and the maximal mixing are strongly disfavoured.

\subsection{Mass classification schemes}

The previously quoted values can be accommodated in the framework of three neutrino mixing, which describes 
the three flavour neutrinos ($\nu_{e}$, $\nu_{\mu}$ and $\nu_{\tau}$) as unitary linear combinations of the three 
massive neutrinos ($\nu_{1}$,  $\nu_{2}$ and $\nu_{3}$) having masses $m_{1}$, $m_{2}$, and $m_{3}$, respectively. 
The best-fit oscillation parameters are summarised in Table \ref{table best fit parameters}.
\begin{table*}
\centering
\caption[]
{Summary of the best-fit values for atmospheric, long baseline accelerator, solar neutrino and long baseline 
anti-neutrino reactor experiments at the begin of Summer 2004.}
\label{table best fit parameters}
\begin{tabular}{cc}\hline
\\
$\Delta{m}^{2}_{23} \simeq 2.4 \cdot 10^{-3} (eV)^2$ & $ \sin^2 2\vartheta_{23} > 0.90 $  for $\vartheta_{13}$=0\\
$\Delta{m}^{2}_{12} = 8.2\pm 0.3 \cdot 10^{-5} (eV)^2$  & $ \tan^2 \vartheta_{12} = 0.39^{+0.05}_{-0.04} $ 
for $\vartheta_{13}$=0\\
$\sin^2 \vartheta_{13} < 0.015 $ \\ 
\hline
\end{tabular}
\end{table*}

The $\Delta{m}^2$ values deduced by the experiments give the following constraint:

\begin{equation}
\Delta{m}^2_{12} \ll
\Delta{m}^2_{23} 
\label{Sun less than atm}
\end{equation}

The experimental measurements are compatible with three mass schemes:
\begin{enumerate} 
\item Normal hierarchy: $m_1 < m_2 < m_3$
\begin{equation}
\Delta\, m^2_{12} \simeq \Delta\,m^2_{\mathrm {Sun}} \simeq m^2_2\qquad
\Delta\,m^2_{23} \simeq \Delta\,m^2_{\mathrm {atm}} \simeq m^2_3
\label{normalhierarchy}
\end{equation}
\item Inverted hierarchy: $m_3 < m_1 < m_2$
\begin{equation}
\Delta\, m^2_{12} \simeq \Delta\,m^2_{\mathrm {Sun}} \qquad
\Delta\,m^2_{23} \simeq - \Delta\,m^2_{\mathrm {atm}} \qquad 
\Delta\,m^2_{13} \simeq m^2_1 
\label{invertedhierarchy}
\end{equation}
\item Degenerate case: the values of $\Delta\, m^2_{ij}$ are small when compared to each mass values. Usually, 
$m_1$ is assumed to be the smaller mass, and the hierarchies are indistinguishable.
\begin{equation}
\Delta{m}^2_{ij} \ll m^2_{1} 
\label{degenerate}
\end{equation}
\end{enumerate}

\begin{figure}[hbt]
\centering
    \begin{minipage}[c]{0.5\textwidth}%
      \includegraphics[height=6.0 cm, angle=0]{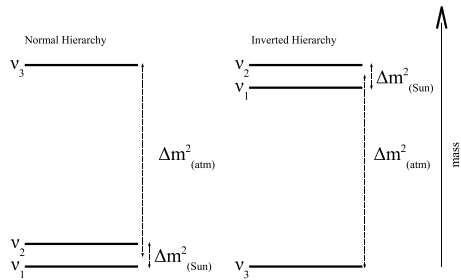}%
    \end{minipage}%
\caption{Neutrino mass schemes based on the experimental relation $\Delta{m}^2_{\mathrm{Sun}} \ll
~\Delta{m}^2_{\mathrm{atm}}$. Presently available experimental data does not allow the identification of 
the right solution.}
\label{hierarchy}
\end{figure}

In figure \ref{hierarchy} "normal" and "inverted" hierarchies are shown. 

At high values of neutrino mass (but below 1 eV), the mass spectrum is practically degenerate. On the contrary, below 
$\sim$ 0.05 eV the degenerate interval splits into two branches: $m_1$ is the lightest mass eigenstate in the normal hierarchy, 
whereas the inverted hierarchy occurs if $m_3$ is the lightest one.

In the normal scheme,  large cancellations are possible between $\nu_{1}$, $\nu_{2}$, and 
$\nu_{3}$ mass contributions independently on the CP conservation, see \cite{Giunti2003};
thus, the $|\langle{m}\rangle|$ value can be arbitrarily small. Then, the smallest 
$\Delta{m}^2$ is realized by the two lightest neutrinos, and a natural neutrino mass hierarchy can be realized 
if $m_1 < m_2$.
 
On the contrary, in the inverted scheme, the cancellations are limited, because $\nu_1$ and $\nu_2$ 
(for $\nu_1$ and $\nu_2$ the electron neutrino has large mixing) are almost degenerate, and much heavier than $\nu_3$, 
independently on its value. The smallest $\Delta{m}^2$ is obtained by the two heaviest neutrinos. 

The elements of the PMNS mixing matrix can be related to the effective mixing angles deduced from experiments. 
Usually, an allowed range for the mixing matrix elements is obtained. Owing of $U_{e3}$ smallness 
($\sim 5 \cdot 10^{-2}$), mainly due to CHOOZ negative results, solar and atmospheric neutrino oscillations are practically 
decoupled, \cite{Bilenky1998}. 

A complete cancellation between the contributions of $\nu_1$ and $\nu_2$ is excluded because the solar 
mixing angle is less than maximal. In the inverted hierarchy, this features puts a lower bound on the effective 
neutrino mass $|\langle{m}\rangle| \simeq$ 0.001 eV. We emphasize that if the mass is smaller than this value,  
either neutrinos have a mass hierarchy or they are Dirac particles. Unfortunately, new $0\nu$-DBD experiments
planned for the next decade will have a sensitivity not better than 0.01 eV, therefore even new detectors
cannot analyse this mass region.

Thanks to the oscillation experiment results, limits on neutrino mass are available:
\begin{itemize}
\item $m_1 \ll m_2 \ll m_3$ \\
In the normal hierarchy case, we obtain:
\begin{equation}
m_1 \ll \sqrt{\Delta\, m_{\mathrm {Sun}}^{2}} \,; \; m_2 \simeq \sqrt{m^2_1 + \Delta\, m_{\mathrm {Sun}}^{2}}\,;
\; m_3 \simeq \sqrt{m^2_1 + \Delta\, m_{\mathrm {atm}}^{2}}
\end{equation}
The presently deduced upper limit to the effective neutrino mass is of $\sim$ 0.0046 eV.
\item $m_3 \ll m_1 < m_2$ \\
In the inverted hierarchy case:
\begin{equation}
m_3 \ll \sqrt{\Delta\, m_{\mathrm {atm}}^{2}} \,; \; m_1 \simeq \sqrt{m^2_3 + \Delta\, m_{\mathrm {atm}}^{2}} \,;
\; m_2 \simeq \sqrt{m^2_3 + \Delta\, m_{\mathrm {atm}}^{2}}
\end{equation}
If the $0\nu$-DBD half-life will be precisely measured, precious information about Majorana CP phase difference 
will be available.
\item $m_1 \simeq m_2 \simeq m_3$ \\
If the effective Majorana neutrino mass is large ($\gg$ 0.045 eV $\sim \sqrt{\Delta\, m_{\mathrm {atm}}^{2}}$), the neutrino 
mass spectrum is almost degenerate. A description in term of $m_1$ is allowed, 
$0.4\,m_1 \leq \left|\langle m_{\nu}\rangle\right|\leq m_1$. If $m_1$ will be determined from beta decay experiments or
from cosmological measurements, accurate value of $0\nu$-DBD half-life could offer strong constraints on Majorana CP phase
differences. 
\end{itemize}

The present best limit on neutrino mass from tritium beta decay experiments is of 2.2 eV, whereas the best value 
from $0\nu$-DBD reactions has been obtained by the Heidelberg-Moscow and IGEX $^{76}$Ge experiments. The upper limit 
ranges from 0.3 to 1.3 eV, depending on the nuclear matrix element value.

In the last two years, new cosmological measurements have put stronger constraints on neutrino mass. Different upper limits 
on the sum of three neutrino masses, at 95\% of C.L., have been computed within the range 0.69 eV, \cite{Spergel2003}, 
and 1.7 eV, \cite{Tegmark2003}, depending on the selected values of associated parameters. This implies an upper limit
to the neutrino mass, which should be not greater than 0.6 eV.

Data from the PLANCK satellite and the SDSS experiment could improve the present sensitivity down to 0.04 eV, but all 
these values are strongly analysis dependent. Taking into account the upper cosmological bound, the absolute scale 
of neutrino mass lies between 0.04 eV and 0.4 eV. Unfortunately, from a theoretical point of view, this is a wide
range of mass, in fact one order of magnitude mass interval allows contradictory conclusions, \cite{Smirnov2004}.

\subsection{Decay processes}

Double beta decays are reactions in which the parent nuclei emit 2 electrons (or positrons) and other light particles. 
At the end of such processes, the decaying nuclei vary their electric charge by two units, but their atomic mass
remains unchanged. Such decays can be observed only if similar processes are absent, i.e. if the intermediate 
nucleus has a mass larger than that of the parent one, or the usual beta decay is strongly forbidden.

Both neutrons and protons can originate DBD reactions, but, in this paper, only the emission of 2 electrons 
is considered, this implies that neutrons transform into protons. 

There are different decaying ways, with or without neutrinos and/or other particle emission. 
In the first one, 2 neutrons (which are bounded in a nucleus) transform 
into 2 protons, and 2 electrons and 2 neutrinos are emitted:
\begin{equation}
(Z,A)  \rightarrow (Z+2,A) + e_1^- + e_2^- + \bar{\nu}_{e_1}  + \bar{\nu}_{e_2}
\label{e:2nu}
\end{equation}
This reaction, which is labelled as $2\nu$-DBD, does not vary the total lepton number L, 
and is fully consistent with the Standard Model of electroweak interactions (SM).
The $2\nu$-DBD half-life is proportional to the Fermi coupling constant, more precisely $\propto G_{F}^{-4}$, and, 
consequently, it is a slow process, \cite{Moe and Vogel 94, Faessler and Simkovic 98}.

An important theorem valid for any gauge model with spontaneous broken symmetry at weak scale states that a $0\nu$-DBD 
process amplitude different from the null value is equivalent to a non-zero Majorana neutrino mass. The same result has 
been extended to the SUSY-versions by \cite{Hirsch1996b}.
The simplest mechanism allowing $0\nu$-DBD reactions is based on left-handed V-A weak currents with the exchange 
of light massive Majorana neutrinos, see \cite{Schechter and Valle1982}; 
in such a reaction two electrons are emitted without neutrinos:
\begin{equation}
(Z,A)  \rightarrow (Z+2,A) + e_1^- + e_2^-
\label{e:0nu}
\end{equation}
This decay was proposed by Racah in 1937, and Furry in 1939, and foresees the violation by two units 
of the total lepton number. It can occurs if and only if neutrino is a Majorana particle (i.e.
the particle is coincident with its own anti-particle). Such a neutrino can be emitted by the first 
neutron decay; then, it reverses its helicity from right to left handed, due to its mass and/or 
a right handed current mixing in weak interactions. Practically, the neutrino
emitted by the first neutron is reabsorbed by the second neutron in the reaction
\begin{equation}
n + \nu_{e}  \rightarrow p^+ + e^-
\label{e:neuabsorption}
\end{equation}

In a $0\nu$-DBD process, only real electrons are created, while two nucleons exchange a 
virtual neutrino. Thus, the electron would carry the decay energy, this is an evident signature of the reaction, 
and makes up for the low probability of this process. The main problem is whether virtual 
neutrino can be exchanged between two identical weak vertices. It is equivalent to the question of
whether a real neutrino can be captured by protons or neutrons, which has been excluded by Davis's experiment 
in 50's, and is reflected in the conservation of lepton number.

Other processes, which violates the total lepton number, may produce $0\nu$-DBD reactions. Among them, we mention 
leptoquarks, supersymmetric particles, and heavy Majorana neutrinos, see \cite{Faessler and Simkovic 98, Hirsch1996b, 
Suhonen and Civitarese 1998, Simkovic and Faessler 2002}. 
The possibility of connections with the equivalence principle and the 
dark energy sector has also been analysed.
The development of grand unified theories (GUTs), such as in the simplest case of $SO(10)$, \cite{Mohapatra1992}, 
left-right symmetric models, or in the minimal supersymmetric standard model (MSSM), extended the electroweak 
$SU(2)_{L}\otimes U(1)$ theories and greatly enhanced the interest in neutrino sector, offering several mechanisms to
allow the $0\nu$-DBD process. 

A third decay mode is the emission of two electrons with Majorons ($\chi$), which are light neutral 
Nambu-Goldstone bosons, due to the spontaneous breaking of a global symmetry associated with the lepton number 
conservation. Such hypothetical neutral pseudoscalar massless particles should be coupled with the neutrino, and 
emitted in the $0\nu$-DBD process,\cite{Berezhiani1992,Hirsch1996}:
\begin{equation}
(Z,A)  \rightarrow (Z+2,A) + e_1^- + e_2^- + \chi (+ \chi)
\label{e:0nuMajoron}
\end{equation}

\begin{figure}[hbt]
\centering
    \begin{minipage}[c]{0.5\textwidth}%
      \includegraphics[height=6.5 cm]{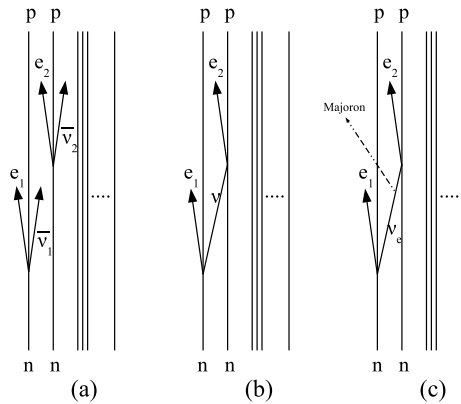}%
    \end{minipage}%
\caption{Feynman diagrams for three DBD cases: a) $2\nu$-DBD process, two anti-neutrinos are present in the final state; 
b) $0\nu$-DBD process, an anti-neutrino is emitted, and then absorbed as a neutrino, the final state is neutrinoless; 
c) DBD process with a Majoron emission.}
\label{Feynman diagrams}
\end{figure}

Some models proposed either the emission of two Majorons (within super-symmetric theories), and a vector Majoron (a longitudinal component of a massive gauge boson). It has been pointed out the importance of the Majoron in the evolution of the early Universe and of the stars.
  
In short, $0\nu$-DBD process can be mediated by the exchange of Majorana neutrinos, light or heavy, but massive, or 
less conventional particles (see figure \ref{Feynman diagrams}). 
Its amplitude, which is strictly related to the mass and coupling costants of such 
particles, is used to check and constrain the adopted parameters.

\subsection{Half-life of DBD processes}

The value of the half-life and the energy value of a physical process give an indication of the 
difficulties experimentalists have to face in order to measure such reaction.

In the case of $2\nu$-DBD process, the inverse half-life T$_{1/2}^{2\nu}$ is free of unknown parameters, 
and depends on exactly calculated integrated phase space factor G$^{2\nu}$,
and $2\nu$-DBD nuclear matrix element, \cite{Doi Kotani and Takasugi 1985}.
\begin{equation}
\left( T_{1/2}^{2\nu}\right) ^{-1}=
G^{2\nu}\cdot
\left|M^{2\nu}_{GT}\right|^2
\label{e:period2nutrinos}
\end{equation}
Because of the large energy release, the most favoured processes are the transitions from the ground state $0^{+}$ 
of the parent nuclei to the ground state $0^{+}$ of the final nuclei.

The $2\nu$-DBD process has been measured for several nuclei, and the obtained half-lives (which are summarised in 
Table \ref{table 2neutrino exp}) vary from $10^{19}$ up to $10^{24}$ y, also representing the weakest measured 
physical process. We remind that in 1987 the experimental discovery of such a process in $^{82}$Se, based on 
non-geo-chemical measurements, was firstly announced, see \cite{Elliott1987} and reference therein.

\begin{table*}
\centering
\caption[]{Summary of experimentally measured half-lives for $2\nu$-DBD. Limits on $0\nu$-DBD reactions involving a standard Majoron are based on \cite{Barabash2004}.}
\label{table 2neutrino exp}
\begin{tabular}{cccc}  \hline
Isotope                & T$_{1/2}^{2\nu}$ (y)           & References & T$_{1/2}^{0\nu \chi}$ (y)           \\  \hline
$^{48}$Ca              & $(4.2 \pm 1.2)\cdot 10^{19}$          & \cite{Balysh 1996,Brudanin 2000}  & $>7.2\cdot 10^{20}$      \\
$^{76}$Ge              & $(1.3 \pm 0.1)\cdot 10^{21}$          & \cite{Klapdor2001,Avignone1991,Aalseth1996}  & $>6.4\cdot 10^{22}$ \\
$^{82}$Se              & $(9.2 \pm 1.0) \cdot 10^{19}$         & \cite{Elliott1992,Arnold98} & $>2.4\cdot 10^{21}$              \\
$^{96}$Zr              & $(1.4^{+3.5}_{-0.5})\cdot 10^{19}$    & \cite{Arnold99,Kawashima1993,Wieser2001} & $>3.9\cdot 10^{20}$ \\
$^{100}$Mo             & $(8.0 \pm 0.6)\cdot 10^{18}$          & \cite{Dassie1995,Ejiri91a,Ejiri91c} & $>5.8\cdot 10^{21}$ \\
                      &                                       &\cite{De Silva97,Alston97,Ashitkov2001,Vasil'ev90} & \\
$^{116}$Cd             & $(3.2 \pm 0.3)\cdot 10^{19}$          & \cite{Arnold96,Danevich00,Ejiri95} & $>3.7\cdot 10^{21}$ \\
$^{128}$Te             & $(7.2\pm 0.3)\cdot 10^{24}$           & \cite{Bernatowicz93,da Cruz93} & $>2.0\cdot 10^{24}$               \\
$^{130}$Te             & $(2.7\pm 0.1)\cdot 10^{21}$           & \cite{Bernatowicz93} & $>3.1\cdot 10^{21}$                     \\
                  & $(6.1\pm 1.4 ^{+2.9}_{-3.5})\cdot 10^{20}$& \cite{Arnaboldi2003}                &     \\
$^{136}$Xe             & $>1.0\cdot 10^{22}$ (90\% C.L.)         & \cite{Bernabei2002} & $>7.2\cdot 10^{21}$                    \\
$^{150}$Nd             & $7.0_{-0.3}^{+11.8} \cdot$ 10$^{18}$  & \cite{De Silva97,Artemiev1995}     & $>2.8\cdot 10^{20}$           \\
$^{238}$U              & $(2.0\pm    0.6)\cdot 10^{21}$        &\cite{Turkevich91}  & --           \\ 
\hline
\end{tabular}
\end{table*}

The nuclear mass element calculation is a very critical point, due to the uncertainties in the description. 
Therefore, experimental measurements of $2\nu$-DBD half-lives can offer strong constraints to the value of 
related nuclear matrix elements, providing important tests of nuclear structure calculations 
\cite{Suhonen and Civitarese 1998}.

The decay probability for the $0\nu$-DBD can be written, see \cite{Doi Kotani and Takasugi 1985,
Suhonen and Civitarese 1998}:

\begin{eqnarray}
\label{decay probability for 0 neutrino decay}
\left( {{\rm T}_{{\raise0.5ex\hbox{$\scriptstyle {\rm 1}$}
\kern-0.1em/\kern-0.15em \lower0.25ex\hbox{$\scriptstyle {\rm
2}$}}}^{{\rm 0}\nu } } \right)^{ - 1}  & = & C_{mm}^{{\rm 0}\nu }
\left( {\frac{{\left\langle {m_\nu  } \right\rangle }}{{m_e }}}
\right)^2  + \;C_{m\lambda }^{{\rm 0}\nu } \left\langle \lambda
\right\rangle \left( {\frac{{\left\langle {m_\nu  } \right\rangle
}}{{m_e }}} \right) +\\ & + & \nonumber
\;C_{m\eta }^{{\rm 0}\nu } \left\langle \eta
\right\rangle \left( {\frac{{\left\langle {m_\nu } \right\rangle
}}{{m_e }}} \right) + \; C_{\lambda \lambda }^{{\rm 0}\nu }
\left\langle \lambda \right\rangle ^2  + \;C_{\eta \eta }^{{\rm
0}\nu } \left\langle \eta  \right\rangle ^2  + \;C_{\lambda \eta
}^{{\rm 0}\nu } \left\langle \lambda \right\rangle \left\langle
\eta \right\rangle 
\end{eqnarray}

In the equation, $\langle m_{\nu} \rangle$, see equation (\ref{neutrino mass}), is the effective
neutrino mass, $m_{e}$ is the electron mass, while $\langle \lambda \rangle$ and $\langle \eta \rangle$ represent 
the effective weak coupling constant of right handed and the left handed nucleonic current. 
For a definition of the $C_{ij}^{0\nu}$ through the specific nuclear matrix elements, and phase space factors 
of $0\nu$-DBD, see for instance \cite{Doi Kotani and Takasugi 1985,Suhonen and Civitarese 1998}.

If all $C_{ij}^{0\nu}$ coefficients are known, all the nuclear matrix element values can be calculated; in this case,
for a given value (or limit) of the $0\nu$-DBD half-life, equation (\ref{decay probability for 0 neutrino decay}) 
represents an ellipsoid which restricts the allowed range of unknown parameters
$0\nu$-DBD: $\langle m_{\nu} \rangle$, $\langle \lambda \rangle$ and $\langle \eta \rangle$ \cite{Moe and Vogel 94}.

If right handed contributions (i.e. $\langle \lambda \rangle = 0 $ and $\langle \eta \rangle = 0$) are not 
taken into account, the half-life of the $0^+ \rightarrow 0^+$ transition can be expressed as for for $2\nu$-DBD reaction, 
see equation \ref{e:period2nutrinos}:
\begin{equation}
\left( T_{1/2}^{0\nu }\right) ^{-1}=G_{mm}^{0\nu }\cdot 
\left|M^{0\nu}_{GT} - \frac{g^2_V}{g^2_A}M^{0\nu}_{F} \right|^2 \cdot 
\langle m_{\nu} \rangle^{2}
\label{e:0nusimple}
\end{equation}
where $m_{\nu}$ is the effective neutrino mass, whereas the $G_{mm}^{0\nu}$ the phase-space integral.  
The nuclear mass elements $M^{0\nu}_{GT}$ and $M^{0\nu}_{F}$ form the nuclear structure parameter $F_N$:
\begin{equation}
F_N = G_{mm}^{0\nu }\cdot 
\left|M^{0\nu}_{GT} - \frac{g^2_V}{g^2_A}M^{0\nu}_{F} \right|^2
\label{e:nuclearparameter}
\end{equation}
In such a transition, an electron is emitted in each vertice of the diagram, and the amplitude of the $0\nu$-DBD 
reaction contains a term $U^2_{ei}$, and is proportional to 
\begin{equation}
\langle m_{\nu} \rangle = 
\left|\sum_i m_i U^2_{ei}\right| 
\label{e:0numassterm}
\end{equation}
where the sum includes only light massive neutrinos. As previously mentioned, the so computed mass is the effective 
neutrino mass, which depends on the Majorana phases, because of the term $U^2_{ei}$ instead $\left|U_{ei}\right|^2$.  
In short, the $0\nu$-DBD rate is directly related to the square of the effective Majorana mass $\langle
m_{\nu} \rangle$, to a calculable phase space factor $G_{mm}^{0\nu }$,
and to the square of a, difficult to compute, nuclear matrix element.

If also Majorons are emitted, the corresponding $0\nu$-DBD rate can be obtained from the previous equation
substituting the effective neutrino mass $\langle m_{\nu} \rangle$ with the effective Majoron neutrino
coupling constant $\langle g_{\chi} \rangle$, and replacing the $G_{mm}^{0\nu}$ factor by the phase space 
integral describing the massless Majoron and the two electrons in the final state \cite{Moe and Vogel 94}:

\begin{equation}
\left( T_{1/2}^{0\nu }\right) ^{-1}=G_{\chi}^{0\nu }\cdot 
\left|M^{0\nu}_{GT} - \frac{g^2_V}{g^2_A}M^{0\nu}_{F} \right|^2 \cdot 
\langle g_{\chi} \rangle^{2}
\label{e:periodMajorons}
\end{equation}

The phase space integrals $G^{2\nu}$, $G_{mm}^{0\nu}$ and $G_{\chi}^{0\nu}$ (where $G = G(Q_{\beta\beta},Z)$) 
contain the Fermi function F$(Q_{\beta\beta},Z)$ which represent the Coulomb distortion of the wave function of 
the emitted electrons. Tabulated values of $G^{2\nu}$, $G_{mm}^{0\nu}$ and $G_{\chi}^{0\nu}$ are collected in 
reviews, see for example \cite{Doi Kotani and Takasugi 1985, Suhonen and Civitarese 1998}.

DBD process toward final excited nuclear states offer several interesting opportunities. The $2\nu$-DBD transition of $^{100}$Mo toward $0^+$ excited state of $^{100}$Ru was successfully detected , see for instance \cite{Barabash2004}, and other isotopes should present half-lives values within actually detectable time intervals ( $\sim 10^{21} - 10^{22}$ y). This suggests that the suppression factor of such reactions is not as great as early supposed and should allow the evaluation of additional properties of nuclear matrix elements.

Neutrinoless DBD reaction is one of the few important non-accelerator experiments which may
demonstrate the validity of GUTs well beyond the possibilities of present and future accelerators.
 
For this reason, since 1948 several collaborations (at present, about 40 groups) 
looked for this rare nuclear process. However, this great effort is till now without results. We remind the first experimental limit to the $0\nu$-DBD process, at a level of  $>3.0 \cdot 10^{15}$ y, based on Geiger detector technique.

Experimentally speaking, if electron energies are well measured, it should be easy to identify such reactions among the 
different decay modes previously quoted. In fact, the energy spectrum of emitted electrons is 
constrained by the phase space of out-coming leptons, and strictly related to the decay process. 
Only indirect methods, as explained in a later section, which allow only the total decay rate measurements, cannot identify 
the decay reaction. The experimental half-life limits allow us to deduce the values for several parameters; 
for example the effective neutrino mass parameters of right handed currents and parameters
of supersymmetric models. 

In a $2\nu$-DBD process the available energy Q$_{\beta\beta}$ is shared between four particles giving origin to a 
continuous spectrum. On the other hand, in $0\nu$-DBD reaction the two electrons
carry the full available energy, giving a sharp peak spectrum at the Q$_{\beta\beta}$ value. 
If Majorons are emitted in a DBD process, the two electron energy spectrum is continuous, but its shape differs from that 
one of $2\nu$-DBD reaction. In fact, the maximum occurs at different energy value, (see figure \ref{Schematic sum 
electron energy spectra}).
We remind that these particles do not interact with ordinary matter and escape detection. In case of transition toward final excited states, an evident experimental signature should be the comparison of one or two photons accompanying the two electrons of fixed total energy.

\begin{figure}[hbt]
\centering
    \begin{minipage}[c]{0.5\textwidth}%
      \includegraphics[height=6.5 cm]{spectrum.eps}%
    \end{minipage}%
\caption{Schematic energy spectra for the emitted electrons calculated for different DBD processes. Each spectrum 
is normalized arbitrarily and independently on the others. In abscissa: the
ratio $E/E_{max}$ between the sum electron kinetic energy divided
by its maximum value.}
\label{Schematic sum electron energy spectra}
\end{figure}

\subsection{The calculation of nuclear matrix elements} 

In order to correctly interpret the $0\nu$-DBD experimental results, the mechanism of nuclear transitions must be 
understood. In other words, one has to evaluate the corresponding nuclear matrix elements with high reliability. 
A fine tuned nuclear matrix calculation, which is based on QCD, is a difficult task with nuclei having several nucleons:
\begin{description}
\item [$\cdot$] The parent nuclei have a complicated nuclear structure, and a many-body approximation in solving 
the calculation is naturally introduced.
\item [$\cdot$] The complete set of states for the intermediate nucleus is a second order weak interaction process.
\item [$\cdot$] There are many parameters involved in the calculations, like pairing interactions, nuclear deformations, 
mean field parameters and so on, and many values have to be fixed. Therefore, the introduced uncertainty is quite high.
\end{description}  
The derived global uncertainty forbids any precise answer to the questions about neutrinos. 

Two are the main approaches to calculate the DBD nuclear matrix elements: the shell model, see 
\cite{Caurier 1996}, and the neutron-proton Quasiparticle Random Phase Approximation (QRPA) model, 
see \cite{Faessler and Simkovic 98, Suhonen and Civitarese 1998}.\\ 
The former was adapted to the DBD sector in the early 80's by \cite{Haxton and Stephenson 84}. 
Improvements were realised mainly for $^{76}$Ge and $^{136}$Xe isotopes. It well describes only the reduced energy 
region of the lowest states for intermediate nuclei, and it does not include the effects from the Gamow-Teller 
resonance region.\\ 
The latter is the most used approach in recent years. It was firstly 
employed in the early 70's, but many problems occurred in reproducing the decay rates for $2\nu$-DBD processes before 
the improvement done by \cite{Vogel and Zirnbauer 1986}. Several upgrades on such model were developed starting from 
the 90's, when DBD physics became more popular. Among them, we mention the Renormalized QRPA, \cite{Toivanen1995, 
Schwieger1996}, the QRPA with proton-neutron pairing, \cite{Cheoun1993}, the full QRPA, \cite{Schwieger1996,Pantis1997}, 
the proton-neutron self consistent RQRPA, \cite{Bobyk1999, Bobyk2001}, and the deformed QRPA, \cite{Simkovic2004}. 
 
A comparison between theories and experiments for the $2\nu$-DBD process provides a measure of confidence in the calculated 
nuclear wave-functions employed for extracting the unknown parameters from $0\nu$-DBD life measurements. 
Usually, the $2\nu$-DBD rates have been estimated not to be fundamental constraints in calculations of $0\nu$-DBD
nuclear matrix elements, because the intermediate nuclear states are quite different. Recently, it has 
been shown that, within the context of QRPA treatment, an accurate knowledge of $2\nu$-DBD rate allows the calculation of 
the nuclear matrix elements reproducing the experimental value for some isotopes, \cite{Rodin2003}.  
Therefore, the previously obtained variability is practically eliminated, and a strong improvement in the corresponding
$0\nu$-DBD nuclear matrix elements seems to be possible.

We stress the great variability which occurs in present calculations. In the case of $^{76}$Ge isotope, the nuclear 
contribution to the decay rate obtained with 20 different approaches is dispersed over about two order of magnitude, 
see for instance \cite{Civitarese and Suhonen 2003, 
Bahcall2004}. This is a consequence of the use in calculation of different methods, model spaces, fitted observables, and adjusted
parametres. Therefore, results also concerning $0\nu$-DBD rates and neutrino mass have a large uncertainty. 

\section{Detection techniques}

\subsection{Preliminary aspects}

In order to have an excellent experiment for DBD searches, which should take place very deep underground, a lot of 
properties should be verified:
\begin{itemize}
\item Large mass detector with compact dimension, looking for the detection of mass value lower than 0.1 eV.
\item The possibility of event reconstruction, with a very good energy resolution.
\item Good source radiopurity, reliable and easy to operate technology.
\item Great natural abundance of the selected decaying isotope, which should have a large Q value, in order to reduce
the background influence.
\item Particle identification, mainly daughter nucleus, and reduced $2\nu$-DBD process interference.
\item Well understood nuclear calculations.
\end{itemize}

Practically, the ideal detector should have excellent sensitivity, in order to identify all the $0\nu$-DBD reactions, 
and almost null background. Clearly, this is a dream. Moreover, all previous requirements cannot be contemporaneously 
satisfied. 

\subsection{Measurement methods}

Since the 50's, the existence of DBD processes was established in pioneering experiments with geo-chemical measurements by searching for daughter nuclei in samples of materials enriched in parent isotopes. In 1949 the $2\nu$-DBD reaction was identified with $^{130}$Te, but only in 1967 another isotope ($^{82}$Se) confirmed such a transition. In fact, when the final 
isotopes are radioactive, also radiochemical measurements can be done. Only the total decay rate can be estimated in this way, but there is a great advantage due to the integration time over very long time. At the end of 80's, the $2\nu$-DBD process was successfully detected in $^{82}$Se by using a Time Projection Chamber, see \cite{Barabash2004} for details and references. 

Up to now, there are two main experimental approaches: 
\begin{itemize}
    \item {\bf{Indirect}}(or Inclusive) = These techniques, which had an important role in the past, 
measure the anomalous concentration of daughter nuclei in samples with a long accumulation time. 
They cannot distinguish between neutrino and neutrinoless processes, and have been used  
to give indirect evaluations of the $0\nu$-DBD and $2\nu$-DBD lifetimes. Among them, we mention geo-chemical and 
radiochemical methods. 
    \item {\bf{Direct}}(or Counter) = These are the presently most diffuse techniques, and are based on the direct 
observation of electrons emitted in the process. Unlike the inclusive method, and according to the different 
capabilities of the detector, energy, momentum, and topology of the decaying particles are recorded in this case. 
These detectors can identify the DBD reaction modes: $0\nu$-DBD process should be easily identified because 
of a mono-energetic line at the Q value. The better the detector energy resolution, the stronger the signal.
Two technical approaches are possible: 
\begin{itemize}
    \item Passive source = The source does not coincide with the detector and hence the electrons 
        are originated in an external sample containing the decaying isotopes.
    \item Active source = The DBD source also serves as the detector.
\end{itemize}
\end{itemize}
 
Different direct standard techniques are employed in DBD experiments, among them:

\begin{itemize}
\item {\bf{Scintillators}}, such as Crystal scintillators and Stacks of plastic scintillators.
\item {\bf{Gas counters}}, such as Time Projection Chambers (TPC) and Multiwire Proportional Chambers (MWPC).
\item {\bf{Solid state detectors}}, such as High Purity Germanium semiconductors detectors (HPGe) and
Silicon detector stacks.
\item {\bf{Bolometers}}
\end{itemize}

\subsection{Background}

In a $0\nu$-DBD experiment, the expected number of reactions strictly depends on the detector mass, and the measurement 
time. On the other hand, the greatest challenge for the experimentalists is the background recognition and elimination
in the energy region around the expected $0\nu$-DBD emission line. 
The way out is to shield the detector and to eliminate internal radioactivity as much as possible.  
Therefore, apparatuses are always located deep underground, in experimental areas like Gran Sasso National Laboratory 
(LNGS) in Italy, Modane in France, Canfranc in Spain, Kamioka in Japan to shield from external components. 
In last years, some new shielded devices in Europe, in Asia, and in America, have opened up new opportunities.

There are many background components interfering with DBD processes detection. Among them, we mention:
\begin{itemize}
\item  Primordial radionuclides, such as $^{238}$U, $^{232}$Th and their daughters, produce dangerous emissions.
Sometimes, their Q-values are as high as the DBD energy region, therefore, a superposition of spectra can occur. 
Further complications are created if a beta decay is promptly followed by an internal nuclear conversion which 
induces a reaction with two electrons. Radon and its daughters also contribute to the background, but their elimination
is frequently obtained by purification in liquid nitrogen. Because of their low Q values, the $^3$H, $^{14}$C, and $^{40}$K
influence should be less dangerous, depending on the selected decaying isotope. 
\item  Anthropogenic (man-made) radionuclides, such as $^{137}$Cs, $^{90}$Sr, $^{42}$Ar, and $^{239}$Pu, which were
produced during atmospheric nuclear tests and/or accidents at nuclear plants.
\item  Cosmogenic isotopes, which have their decaying energy in the $0\nu$-DBD region. The intensity of different 
contributions is material dependent. It is hard to eliminate such components, even if a laboratory located deep underground
greatly reduces their influence. 
\item  Neutrons, due to the difficulty to identify such neutral particles: both neutron capture and fast neutron 
interaction. If the detector is located deep underground, these reactions are reduced, but a good knowledge of the neutron
flux is required.
\item Cosmic ray muon-induced events. A good solution is the deep underground location of the detector combined with a 
veto system to eliminate the prompt interaction via coincidence technique.
\item $2\nu$-DBD reactions. Usually, the $2\nu$-DBD energy spectrum has low intensity in the Q$_{\beta\beta}$ region, but 
it can produce a dangerous background. A great energy resolution is needed in order to well identify true signals; 
sometimes, an asymmetric energy window, centred at Q value, is also selected.     
\end{itemize}

In the last decades, several techniques aiming for a more complete evaluation of the background, and, consequently, 
its consistent reduction have been developed. Examples of this are: pulse shape analysis, topological information of 
the events, simultaneous measurements of two signals such as heat and ionisation, or heat and scintillation and so on. 
Moreover, the strongly reduced width of the peak at the Q$_{\beta \beta }$ value can be masked by the $2\nu$-DBD
tail for some DBD candidate nuclei. For this reason, a very good energy resolution is also necessary
to detect the $0\nu$-DBD signal.

\subsection{Detector performances}

Experimentally speaking, a $0\nu$-DBD detector has to identify a known energy peak within a continuum 
spectrum, where energy lines due to different radioactive isotopes can be also present. Therefore, a wide 
knowledge of the background shape and intensity in surrounding energy region is needed in order to  
analyse the detected emissions.     

Once the detector parameters are fixed, it is possible to calculate the expected number of background events, 
$N_B$, in an energy interval equal to the apparatus FWHM energy resolution, centred around the transition energy:
\begin{equation}
N_B=B \cdot \Delta E \cdot t \cdot m 
\label{NB}
\end{equation}
In case of constant background level $B$, usually expressed in counts/(keV kg y), the background counts linearly scale 
with the measurements time $t$, the sensitive mass of the detector $m$, and the energy resolution $\Delta E$. 
Consequently, the half-life limit of $0\nu$-DBD can be written in the form:
\begin{equation}
T_{1/2}^{0\nu }  \sim \frac{a}{W} \cdot \varepsilon \cdot \sqrt{ \frac{m \cdot t}{B \cdot \Delta E} }
\label{sensitivity simple}
\end{equation}
where $\varepsilon$ is the detection efficiency, $a$ the isotopic abundancy, and $W$ the molecular weight. 
About the detection efficiency, only direct methods allow a complete detection (100\%).

As long as background is null, half-life grows as the sensitive mass and the measurement time, whereas in the case 
of background counts the dependence is on the square root of the same quantities.
We consider two detectors with the same values of efficiencies, resolution, background, and running time, but 
with different masses and isotopic enrichments. The same sensitivity is obtained if the ratio between the masses 
is equal to the square of the inverted ratio between the isotopic abundancy. Usually, the mass under observation is 
isotopically enriched in the selected decaying nucleus, but this is a very expensive process. 
A relatively cheap technique is based on centrifugal isotope separation when the substance is in gaseous form, but it
can be applied only to $^{76}$Ge, $^{82}$Se, $^{100}$Mo, $^{116}$Cd, $^{130}$Te, and $^{136}$Xe isotopes. At present, 
only Russian plants allow this enrichment process. An alternative, but expensive method should use the atomic 
vapour laser isotope separation, even if the production program is up to now not planned. 
In this case, $^{48}$Ca, $^{100}$Mo, $^{116}$Cd and $^{150}$Nd enriched materials could be obtained at 
Livermore National Laboratory, USA. Moreover, a large mass production is possible only for 
some DBD candidate isotopes: e.g. $^{76}$Ge, $^{82}$Se, $^{116}$Cd, $^{130}$Te, and $^{136}$Xe. 

A useful parameter is the detector sensitivity (or detector factor of merit), which is defined as 
the process half-life corresponding to the maximum signal (N$_B$) that could be hidden by the background fluctuations, 
at a given statistical C.L., \cite{Cremonesi2002}. In other words, the sensitivity is the lifetime corresponding to 
the minimum detectable signal above background fluctuations. It allows also the comparison of the performance among 
different experimental apparatuses. 
$F_{0\nu}$, which represents the inverse of the minimum rate detectable in a measurement time t, can be estimated, 
at 1$\sigma$ level:
\begin{equation}
F_{0\nu}  = T_{1/2}^{Backg.}= \ln 2 \cdot
N_{\beta\beta} \cdot \varepsilon \cdot \frac{t}{\sqrt{N_B}} = \ln 2 \cdot (N_A\,k) \frac{a\,\varepsilon}{W} \sqrt{
\frac{m\,t}{B\,\Delta E}}
\label{sensitivity one}
\end{equation}
where $N_{\beta\beta}$ is the number of observed $\beta\beta$ decaying nuclei, $N_A$ is the Avogadro number, and $k$ is the 
number of decaying nuclei per molecule. In this equation the role of each component is clearly emphasized.
By equation (\ref{e:0nusimple}), it is also possible to deduce the experimental sensitivity to the neutrino mass,
$F_{\langle m_{\nu} \rangle}$:
\begin{eqnarray}
F_{\langle m_{\nu} \rangle} = \sqrt{\frac{1}{F^{Exp}_{0\nu} \cdot G_{mm}^{0\nu} \cdot |NME|^2}} = \nonumber \\
= \sqrt{\frac{W}{N_A\,k\,a\,\varepsilon \, G_{mm}^{0\nu}\, |NME|^2}} \cdot 
\left(\frac{B\,\Delta E}{m\,t}\right)^{\frac{1}{4}}
\label{sensitivity on neutrino mass}
\end{eqnarray}
where NME are the nuclear matrix elements.

It is straightforward to conclude that very large sample masses (possibly enriched with DBD candidate nuclei) 
and very low background are needed to look for the identification of the effective neutrino mass. 
We point out that a sensitivity of $\sim$ 0.01 eV is required in order to check inverse hierarchy.

\begin{figure}[hbt]
\centering
    \begin{minipage}[c]{0.5\textwidth}%
      \includegraphics[height=6.5 cm]{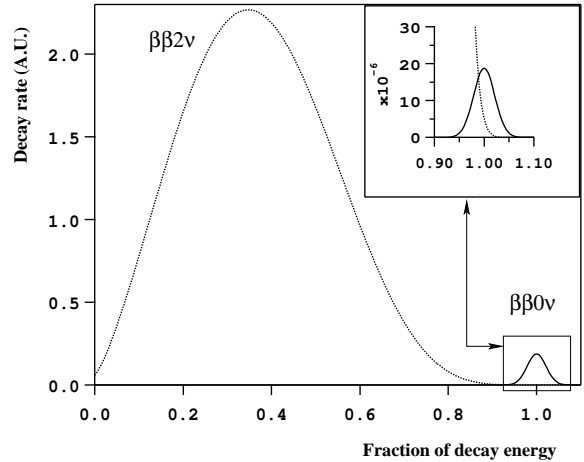}%
    \end{minipage}%
\caption{Energy spectra for the electrons emitted in $2\nu$-DBD (dotted line) and $0\nu$-DBD (full line) reactions, based on an energy resolution of 5\%. On the X axis, the fraction between 
electron kinetic energies and the Q value is represented. The intensities have been normalized to different values for of $2\nu$-DBD and  $0\nu$-DBD terms. In the upper inset, the contribution of $2\nu$-DBD reaction to the background of $0\nu$-DBD reaction has been enhanced, see \cite{Elliott2004}. }
\label{2nu tail}
\end{figure}

Among the components of the background, the $2\nu$-DBD process can produce dangerous emissions. In fact, all the features
of the two decay modes (with and without neutrinos) are equal: two electrons are emitted in one point 
inside the source, at the same time, in the same energy region and with the same angular distribution. 
No available techniques of discrimination can distinguish between $0\nu$-DBD and $2\nu$-DBD 
signals, and consequently, it is impossible the rejection of $2\nu$-DBD contributions. The latest, 
and more energetic, part of the $2\nu$-DBD spectrum overlaps the gaussian peak of the $0\nu$-DBD process. 
Therefore, a very important parameter is the energy resolution of the detector. The better this is, the smaller will 
be the undesirable $2\nu$-DBD contribution to the background in the analysed region (see figure \ref{2nu tail}).

As a further improvement, a very good estimate of the neutrino fluxes is required in experimental areas. We mention the solar neutrinos, namely the high energy component, the anti-neutrinos from nuclear power plants and from the Earth. This contribution produces a background which can significantly influence the instrumental sensitivity.

\section{Experimental status}

In this section we will briefly review some of the direct counting experiments, reporting only on DBD reactions to the ground state. The direct counting experiments using transitions toward excited states will not be detailed in the present paper, see \cite{Barabash2004} for a comprehensive analysis and the present experimental limits quoted in Table \ref{table 2neutrino exp}.

\subsection{Present Results}
\begin{enumerate}

\item $^{48}$Ca is the most favourable isotope among other potential $2\nu$-DBD nuclei because it 
has the largest Q value (4272 keV), hence the possibility of the occurrence is highest,
and the expected background should be lower than in remaining candidate nuclei.
A new CaF2 scintillation detector system (ELEGANTS VI), which consists of 6.6 kg of CaF$_2$(Eu) crystals as 
sensitive mass, has been developed at the Oto Cosmo observatory, 
near Nara in Japan. The obtained energy spectrum after all cuts 
gives a lower limit for the half life of $\rm T_{1/2}^{0\nu} > 1.4 \cdot 10^{22}$ y \cite{Ogawa2004}.

\item Two experiments have looked for the DBD of $^{76}$Ge nucleus. 
\begin{itemize}
    \item[$\cdot$] The Heidelberg-Moscow (HM) experiment, \cite{Klapdor2001}, which is located at LNGS, and
employs a set of five large HPGe detectors enriched in $^{76}$Ge to 86-88\%. The active total mass is $\sim$ 11 kg, 
which corresponds to 125.5 moles of $^{76}$Ge. Due to the passive, consisting of extremely low background materials, 
and active shields, the background is strongly reduced, $\simeq 0.2$ counts/(keV kg y)
in the peak region. After the pulse shape discrimination analysis, its value is lowered to 
$0.113 \pm 0.007$ counts/(keV kg y) in the period 1995-2003, in the $0\nu$-DBD peak energy 
region, where the energy resolution is 3.9 keV.\\
The data taking, which began in 1990 with a reduced setup, stopped in May 2003, for a total value of 71.7 kg$\cdot$y. 
The half-life of the process has been determined on the basis of more than $10^5$ events. 
Some researchers of this collaboration have recently claimed the discovery of the $0\nu$-DBD process at a level 
of 4.2 $\sigma$, see \cite{Klapdor2004} and references therein. We will briefly discuss this results in 
following sections. 
    \item[$\cdot$] The IGEX detector, which is homed at the Spanish laboratory of Canfranc, with a shield of 
about 4000 m.w.e., consists of three HPGe detectors enriched in $^{76}$Ge up to 88\%, with a total active mass 
of at least 6 kg. 
After pulse shape discrimination analysis, the background rate is as great as in HM experiment
in the energy interval between 2.0 and 2.5 MeV, while the energy resolution is 4 keV. 
Analysis on 8.9 kg $\cdot$ y ($^{76}$Ge) of data gives a
lower bound of T$_{1/2}^{0\nu} > 1.57 \cdot 10^{25}$ y \cite{Aalseth1999}. 
\end{itemize}
It has to be stressed that even if both the experiments gave an effective neutrino mass limit of 0.3 - 1.3 eV, 
IGEX detector has a background of 0.01 counts/(keV kg y), mainly internal(cosmogenic), whereas in 
old HM analyses the background was estimated as great as 0.06 counts/(keV kg y), mainly external.
\item Several collaborations have worked on $^{100}$Mo isotope, 
in particular NEMO, in France, and ELEGANT V, in Japan.
\begin{itemize}
\item[$\cdot$] NEMO-3 experiment, which started its data taking in February 2003, 
is homed at Fr\'ejus Underground Laboratory at a depth of $\sim$ 4800 mwe. It is an improvement of NEMO-2 and analyses also the $2\nu$-DBD 
reactions of $^{82}$Se, $^{96}$Zr, $^{100}$Mo, and $^{116}$Cd. It is a cylindrical tracking detector (see Figure \ref{Nemo 3 exp}), divided into 20 equal 
sectors, and devoted to the search for $0\nu$-DBD processes with passive sources enriched up to 97$\%$ in $^{100}$Mo ($\sim$
7 kg). Thin (40-60 mg/cm$^2$) enriched foils of $\beta\beta$ emitters have been constructed from either 
metal films or powder bound by an organic glue to mylar strips \cite{Etienvre2003,Arnold2004}. Its present FWHM at the 
Q$_{\beta\beta}$ values is of 90 keV.
The expected sensitivity for the effective neutrino mass is on the order of 0.2 - 0.3 eV after 5 years of measurements. \\
At present, several interesting results available. In Table \ref{table Nemo3 neu} $2\nu$-DBD main characteristics are shown, whereas $0\nu$-DBD values are quoted in 
Table \ref{table 0neutrino exp}. 

\begin{table*}
\centering
\caption[]{Results from NEMO-3 experiment for $2\nu$-DBD reactions, 
based on more than 140000 detected events \cite{Sarazin2004}. The statistics and the systematic errors are also quoted. Its has to be stressed the great variability in mass among different analysed isotopes. For a comparison with previous values see Table \ref{table 2neutrino exp}.}
\label{table Nemo3 neu}
\begin{tabular}{lcc}  \hline
Isotope    &Mass (g)           & T$_{1/2}^{2\nu}$ (y)  \\ 
\hline

$^{82}$Se   &932 & $ (10.3 \pm 0.3 \pm 0.7) \cdot 10^{19}$   \\ 
$^{96}$Zr   &9.4 & $ (2.0 \pm 0.3 \pm 0.2 ) \cdot 10^{19}$   \\
$^{100}$Mo  &6914 & $ (7.68 \pm 0.02 \pm 0.54) \cdot 10^{18}$   \\
$^{116}$Cd  &405 & $ (2.8 \pm 0.1 \pm 0.3) \cdot 10^{19}$   \\
$^{150}$Nd  &37 & $ (9.7 \pm 0.7 \pm 1.0) \cdot 10^{18}$   \\
\hline

\end{tabular}
\end{table*}

\begin{figure}[hbt]
\centering
    \begin{minipage}[c]{0.5\textwidth}%
      \includegraphics[height=6.5 cm]{nemo3.eps}%
    \end{minipage}%
\caption{Schematic view of the NEMO-3 experimental setup where 1) indicates $\beta\beta$ isotope foils. The energy of electrons is measured by plastic scintillators 2) coupled to low activity PMTs 3). Moreover,  6180 drift cells operating in Geiger mode 4) allow the track resolution with a 1 cn resolution. In addition, a solenoid surrounding the apparatus produces a 25 G magnetic field parallel to the detector axis. The external passive shield (steel, water, wood and paraffin) have been removed.}
\label{Nemo 3 exp}
\end{figure}
 \item[$\cdot$] The most stringent half-life limit is the one obtained 
by the ELEGANT V spectrometer in the Oto Cosmo Observatory by 
Osaka University. The detector consists of three  drift chambers whose aim is to 
detect two $\beta$ trajectories, a sodium iodide crystal scintillator
array to detect $\gamma$ rays, and plastic scintillators to measure the $\beta$  
ray energies and arrival times \cite{Ejiri2001}. The passive source consists of
two foils enriched in $^{100}$Mo up to $95\%$ whose thickness is 20 $\rm mg/cm^{2}$ with a 
total mass of $\sim$ 170 g inserted in a central drift chamber.
The lower limit thus obtained is T$_{1/2}^{0\nu} > 5.5 \cdot 10^{22}$ y \cite{Ejiri2001}.
\end{itemize}

\item Experiments with $^{116}$Cd have been made by the Ukrainian Institute of Nuclear Research INR-Kiev 
(since 1998 in collaboration with the University of Florence) in the salt mine of Solotvina (Ukraina). 
The lower limit is T$_{1/2}^{0\nu} > 1.7 \cdot 10^{23}$ y.

\item $^{130}$Te isotope (and also $^{128}$Te) has been investigated by an Italian group based in Milan (INFN and 
Milan-Bicocca University), within the project MIBETA, by developing low temperature thermal detectors in the 
form of TeO$_2$ crystals (i.e. bolometers). This nuclide was primarily selected because of its natural isotopic 
abundance (34\%), its high transition energy and a favourable nuclear factor of merit. Moreoever, also geo-chemical 
techniques which use this isotope are available from long time.\\
The detector was homed at LNGS, and consists of an array of 20 TeO$_{2}$ crystals, with a total mass of 6.8 kg, 
which operate at a temperature of $\sim$ 12 mK; its resolution is 8 keV in the $0\nu$-DBD region (2528 keV). 
The background level in the same region is  0.33 $\pm$ 0.11 counts/(keV kg y).
The lower limit is T$_{1/2}^{0\nu} > 2.1 \cdot 10^{23}$ y, corresponding to a 
range 0.9 - 2.1 eV in $\langle m_{\nu} \rangle$ \cite{Cremonesi2002}.

Since February 2003, a prototype called CUORICINO, which consists of an array of TeO$_{2}$ bolometers, 
for a total mass of 40.7 kg, is running at LNGS. The array is composed by 2 modules, 9 detectors each with 3 x 3 x 6 cm$^3$ 
crystals, and 11 modules, 4 detector each having 5 x 5 x 5 cm$^3$ crystals. Its FWHM at the Q$_{\beta\beta}$ value is
7 keV, \cite{Giuliani2003, Arnaboldi2004}. The background presently measured is 0.19 $\pm$ 0.02 counts/(keV kg y), and the live time is of 5.8 kg$\cdot$y. No evidence of $0\nu$-DBD reactions has been detected during 2003 data taking. 
A half-life limit of 7.5$\cdot 10^{23}$ y at 90 \% of C.L., which corresponds to a neutrino mass 
interval between 0.3 and 1.6 eV, has been deduced \cite{Fiorini2004}.
The estimated sensitivity after 3 years of data acquisition is 4$\cdot 10^{24}$ y, or, in mass, 0.2 - 0.5 eV, 
It will be an important test of CUORE project feasibility (see later) both for technical performance, and background level 
expectations \cite{Arnaboldi2003}.

\item The $0\nu$-DBD of $^{136}$Xe has been used for the Caltech-Neuchatel-PSI collaboration and the Italian 
group DAMA-LXe.
\begin{itemize}
\item The Caltech-Neuchatel-PSI detector consist of a time projection chamber with a total active
volume of $\sim$ 180 l, containing 3.3 kg of Xe gas ($\simeq$ 24 moles) enriched in $^{136}$Xe to 62.5\%
at a pressure of 5 atm. The detector is located in the Gotthard underground laboratory in the Swiss Alps.
At Q$_{2\beta}$ = 2481 keV, the FWHM energy resolution is 6.6\%. The background rejection is assured by
the time projection chamber track reconstruction, and its value is $\sim$ 0.02
counts/(keV kg y) in the Q$_{2\beta}$ region (within a FWHM interval energy).
The obtained lower limit is T$_{1/2}^{0\nu} > 4.4 \cdot 10^{23}$ y \cite{Luescher1998}.
\item A better result has been obtained by the Roma group at LNGS, by using $\sim$ 6.5 kg of high purity  
liquid Xenon scintillator has been filled by (Kr-free) Xe gas enriched in $^{136}$Xe (68.8\%), and in $^{134}$Xe (17.1\%).
The statistics was 1.1 kg $\cdot$ y for $^{134}$Xe, and 4.5 kg $\cdot$ y for $^{136}$Xe.
The lower limits obtained for half-life were T$_{1/2}^{0\nu} > 1.2 \cdot 10^{24}$ y for 
$^{136}$Xe and T$_{1/2}^{0\nu} > 5.8 \cdot 10^{22}$ y for $^{134}$Xe, at 90\% of C.L., \cite{Bernabei2002}.
\end{itemize}
\end{enumerate}

\begin{table*}
\centering
\caption[]{Experimental 90 \% C.L. half-life limits for $0\nu$-DBD, except where noted. 
The effective neutrino mass upper limits and ranges have been deduced by the authors.
The intermediate rows show the results concerning the claimed discovery of $0\nu$-DBD reaction: 
in upper row the 3$\sigma$ interval is reported, whereas in the lower one the best fit values are quoted. In the final part, the latest published results.}
\label{table 0neutrino exp}
\begin{tabular}{llll}  \hline
Isotope                & T$_{1/2}^{0\nu}$ (y) & References &
$\langle m_{\nu} \rangle$ (eV) \\ 
\hline

$^{48}$Ca    & $> 1.4 \cdot 10^{22}$                     & \cite{Ogawa2004}         &$<7.2-44.7$      \\ 
$^{76}$Ge    & $> 1.9 \cdot 10^{25}$                     & \cite{Klapdor2001}      &$<0.35$     \\ 
$^{82}$Se    & $> 2.7 \cdot 10^{22}$ (68\%)              & \cite{Elliott1992}        &$<5.0$        \\ 
$^{100}$Mo   & $> 5.5 \cdot 10^{22}$                     & \cite{Ejiri2001}         &$<2.1$      \\ 
$^{116}$Cd   & $> 1.7 \cdot 10^{23}$                     & \cite{Danevich2003}   &$<1.7$  \\ 
$^{128}$Te   & $> 7.7 \cdot 10^{24}$                     & \cite{Bernatowicz93}     &$<1.0-4.4$  \\ 
$^{130}$Te   & $> 5.5 \cdot 10^{23}$                     & \cite{Arnaboldi2004}     &$<0.37-1.9$  \\ 
$^{134}$Xe   & $> 5.8 \cdot 10^{22}$                     & \cite{Bernabei2002}      &$<17.0-27.0$  \\ 
$^{136}$Xe   & $> 1.2 \cdot 10^{24}$                     & \cite{Bernabei2002}      &$<0.8-2.4$  \\ 
$^{150}$Nd   & $> 1.2 \cdot 10^{21}$                     & \cite{De Silva97}        &$<3.0$        \\
\hline
$^{76}$Ge    & $(0.69-4.18)\cdot 10^{25}$                &\cite{Klapdor2004}        &$0.24-0.58$     \\ 
$^{76}$Ge    & $1.19 \cdot 10^{25}$                      &\cite{Klapdor2004}        &$0.44$     \\ 
\hline
$^{82}$Se    & $> 1.4 \cdot 10^{23}$                     & \cite{Sarazin2004}        & $<1.5-3.1$  \\ 
$^{100}$Mo   & $> 3.1 \cdot 10^{23}$                     & \cite{Sarazin2004}     & $<0.8-1.2$  \\ 
$^{130}$Te   & $> 7.5 \cdot 10^{23}$                     & \cite{Fiorini2004}       &$<0.3-1.6$  \\ 

\end{tabular}
\end{table*}

Half-life limits have been established experimentally for several nuclides; table \ref{table 0neutrino exp} 
summarizes the measured values for $0\nu$-DBD and the effective neutrino mass $\langle m_{\nu} \rangle$ limits, 
or ranges, as deduced by the authors of different experiments. These results have already put the strongest 
constraint on the Majorana 
neutrino mass, which can vary between 0.3 eV and 5 eV, the right handed admixture in the weak interaction 
($\eta \sim 10^{-7}$ and $\lambda \sim 10^{-5}$), the coupling constant between neutrino and Majoron 
($g_{M} \sim 10^{-4}$), and the R-parity violating parameter in the MSSM ($\zeta \sim 10^{-4}$).

Direct measurements of $2\nu$-DBD gave positive results for several isotopes, the last ones being from $^{76}$Ge, $^{96}$Zr, $^{100}$Mo, $^{116}$Cd, and $^{150}$Nd. The values vary between $10^{19}$ and $10^{21}$ y, see Table \ref{table 2neutrino exp}.

The strongest limits on $0\nu$-DBD half-life, and, consequently, on neutrino mass, come from enriched $^{76}$Ge 
IGEX and HM experiments, which recently stopped their measurements after several years of data taking.
Their results are consistent both in background level, $\sim$ 0.2 counts/(keV kg y), before the pulse 
shape discrimination analysis, and in half-life limit, within the range 1.3 - 1.9 $\cdot 10^{25}$ y.

At the present moment, only NEMO-3 and CUORICINO detectors are running. 

\section{Next generation of DBD experiments}

In this section we will review the future projects in the $0\nu$-DBD research field either in the R\&D phase, 
or simply submitted proposals; they are ordered following the nuclear mass of the analysed isotope.
\begin{enumerate}

\item {\bf{Calcium}} \\
CANDLES (CAlcium fluoride for studies of Neutrino and Dark matters by Low Energy Spectrometer) is based on the 
use of CaF$_{2}$ immersed liquid scintillator at the Oto Cosmo Observatory, in Japan.  
Several steps have been planned: the CANDLES III setup consists of 60 crystals (3.2 kg each), for a total mass of $\sim$ 200 kg, 
and a resolution below 4\% at 4.27 MeV. After 3 years of data taking, the sensitivity on $\langle m_{\nu} \rangle$ 
will be 0.5 eV. The upgraded setup, called CANDLES IV, consisting of 1000 crystals (3.2 kg each) for a total mass 
of $\sim$ 3.2 ton, should reach a $\langle m_{\nu} \rangle$ limit of 0.150 eV.
In the case of $\rm ^{48}Ca$ enrichment from the natural abundance of 0.18\% to 2.0 \% (called CANDLES V), 
the limit on the sensitivity will be $\sim$ 0.030 eV. The same result could be obtained without enrichment, but a total
mass of $\sim$ 50 ton and a low background would be needed \cite{Umehara2003, Kishimoto2004}.

\item 
{\bf{Germanium}} \\
The planned experiments are GENIUS, MAJORANA, and GEM, but in spring 2004 a further project has been presented. For all 
these ionisation detectors, the cooling solution is given by using a cryostat, like in MAJORANA, or a liquid 
Nitrogen bath, in the remaining ones.

\begin{itemize}
\item The GENIUS experiment(GErmanium in liquid NItrogen Underground Setup) would consists of 400 enriched 
(86 - 88\%) HPGe naked crystals, for a total mass of $\sim$ 1 ton. The detector will be immersed in a
liquid nitrogen bath, which also serves as high purity passive shield. To prove the feasibility of 
this detector, three small naked HPGe crystals have been tested in liquid nitrogen. The result is comparable 
to that of conventional HPGe diodes (i.e. in vacuum-tight cryostat). The use of naked crystals should move
the external radioactivity to outside the liquid nitrogen region. 
The quoted energy resolution is $\sim$ 6 keV, while the expected background, which should be maximally due to the 
external component, is $\sim$ 0.0001 counts/(keV kg y), and the estimated sensitivity on mass
is 0.015 - 0.045 eV. A test of a naked crystal operating in a liquid nitrogen filled dewar was successful; therefore, a
prototype (GENIUS - Test Facility), consisting of 14 naked HPGe crystals was already 
approved by the LNGS scientific committee.

\begin{figure}[hbt]
\centering
    \begin{minipage}[c]{0.5\textwidth}%
      \includegraphics[height=6.5 cm]{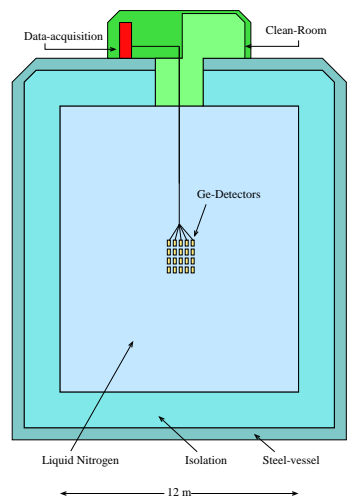}%
    \end{minipage}%
\caption{Proposed experimental setup for GENIUS detector. An array of 1 ton of enriched $^{76}$Ge is hanging on a 
structure in the middle of a tank filled by liquid nitrogen. The size for this apparatus is greater than 12 m. 
A clean room and data acquisition room are on the top.}
\label{GENIUS exp}
\end{figure}

\item The MAJORANA setup will consist of 210 86\%-enriched HPGe crystals (as segmented diodes) for a total mass 
of $\sim$ 0.5 ton, but, unlike GENIUS, very low activity conventional cryostats 
will be employed. Digital electronics and improved pulse shape discrimination will also be used.
The estimated background is $\sim$ 7.3 counts in an energy region of 3.6 keV, which corresponds to
$\sim$ 0.0001 counts/(keV kg y), and a half-life value of $\sim$ $\cdot 10^{27}$y at 90 \% of C.L.
The main component of the background reduction will be the granularity of the detector. Among different aspects under 
analysis, a prototype should check the cooling process for multiple crystals within a single low-background cryostat, 
and also the performance in rejecting the background of a segmented detector configuration. Its energy resolution at the
Q$_{\beta\beta}$ value is 4 keV.
The expected sensitivity for an experimental running time of 10 y is in the 0.030 - 0.040 eV range, 
\cite{Aalseth2004}. The detector is planned to be installed  at Waste Isolation Pilot Plant 
(WIPP), near Carlsbad, in the USA.  

\begin{figure}[hbt]
\centering
    \begin{minipage}[c]{0.5\textwidth}%
      \includegraphics[height=6.5 cm]{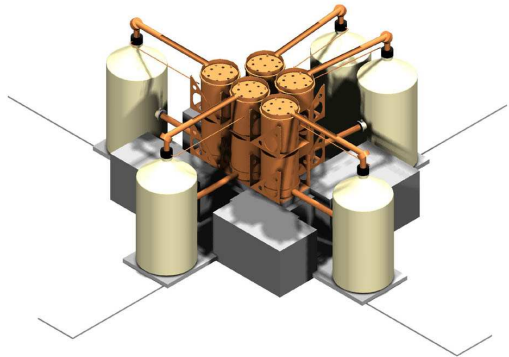}%
    \end{minipage}%
\caption{Schematic view of a possible experimental setup for Majorana detector. The external cylindric tanks are 
liquid nitrogen dewars, whereas inner cylinders are the copper cryostats containing germanium detectors. 
Lead blocks are also shown.}
\label{Majorana exp}
\end{figure}

\item The GEM project should use $\sim$ 1 ton of naked HPGe detectors operating in super-high purity liquid 
nitrogen contained in a copper vacuum cryostat. The detector is within a 5 m diameter sphere placed 
in a water shield. The first GEM-I phase will employ natural germanium, the GEM-II step will be enriched 
in $^{76}$Ge at 86\%, \cite{Zdesenko2001}. 

\item An interesting proposal, which plan to merge, in an unique detector, all the HP$^{76}$Ge elements
previously used by IGEX and HM collaborations, has been presented by some members of HM collaboration and other 
researchers \cite{Abt2004}. This new apparatus, 
which could reach an active mass of more than 20 kg, would reside at LNGS. A configuration with a 1.5 m of liquid 
Nitrogen/liquid Argon shield surrounded by $\sim$ 10 cm of high-purity lead inside the cryostat is under analysis. 
A further external 2-m of water shield should prevent (and identify) contaminations due to rock and concrete, 
neutrons and cosmic rays, if photomultipliers are added. 
If funded, its construction could start in early 2005, whereas its data acquisition could begin in 2006.
After 1 year of data taking it could confirm or refuse with a high level of significance the claimed 
discovery of $0\nu$-DBD reaction by using the same isotope. As a second step, a further 
addition of 20 kg of enriched HP$^{76}$Ge is also planned.
\end{itemize}
\item
{\bf{Selenium}}  \\
SuperNEMO would be an improvement of NEMO-3 detector, with $\sim$ 100 kg of foils enriched in $^{82}$Se, and better energy resolution. A neutrino mass sensitivity in the range 0.04 - 0.15 eV is expected, corresponding to a half-life limit greater than $10^{26}$ y. The proposed apparatus should consist of four sections, each of $\sim$ 2 x 3 x 20 m$^3$, surrounding the very low radioactive mixture of $^{82}$ Se. The electron energy will be measured by plastic scintillators (having an energy resolution of $\sim$ 10\% at E = 1 MeV), whereas Geiger counters will reconstruct the particle tracks. The use of $^{100}$ Mo, $^{116}$Cd and $^{130}$Te isotopes is also possible. A general upgrading of the underground facilities is still required, see \cite{Barabash2004}.
\item
{\bf{Molybdenum}}  \\
In Japan, the MOON (MOlybdenum Observatory of Neutrinos) experiment will use $^{100}$Mo as active target, aiming the detection
of low energy solar neutrinos (at $E>$ 168 keV), and $0\nu$-DBD reactions.  
The detector is sensitive to $0\nu$-DBD via the $^{100}$Mo decay to the ground and excited state of $^{100}$Ru. 
The setup will be a huge sandwich made by foils of natural molybdenum interleaved with a specially designed plastic scintillator. 
The molybdenum total mass will be large, \cite{Ejiri2000d}. High purity levels for the scintillator are needed, and a 
great effort is required in this sector, because of the large surface. Other options, such as 
metal-loaded liquid scintillator and bolometers, have been analysed. The resolution at the Q-value (3.034 MeV) 
should be $\sim$ 7\%. Two intermediate steps have been planned to check the feasibility of the final configuration.
An enrichment process has been also considered in order to reduce 
the detector dimensions and the internal radioactivity, but it is very expensive. 
\item
{\bf{Cadmium}}\\
The planned experiments are:
\begin{itemize}
\item
The COBRA (Cadmium-Telluride O neutrino double Beta Research Apparatus) collaboration also plans 
to use $^{130}$Te candidates (and $^{116}$Cd) under the form of a new generation of semiconductors. These ionisation
detectors, which operate at 300 K with an energy resolution of $\sim$ 1\% at the 661 keV line, are quite 
small in size, and allow systematic studies on Cd and Te isotopes, and rare beta decays 
of $^{113}$Cd and $^{123}$Te. Up to now, only 1 cm$^{3}$ diodes, which corresponds to $\sim$ 6 g, 
have been exploited. The COBRA apparatus is planned to run with an array of $\sim$ 15 x 15 cm, which corresponds to 
a mass of $\sim$ 1.3 kg. The detector can be extended by stacking additional modules to form a 
tower, and by adding more towers later on, \cite{Kiel2003}. 
\item
CAMEO is an upgraded version of the experiment on $^{116}$Cd performed in Solotvina underground laboratory. The initial step
will use 24 enriched cylindrical $^{116}$CdWO$_4$ crystals, with a total mass of 65 kg, and will be placed 
in the middle of the Counting Test Facility (CTF), at LNGS. In order to have the required optical coverage, the present number of 
photomultipliers will be doubled. The total background in the energy region of interest has been estimated to be
$\sim$ 3 counts/y. After a measuring time of more than 5 years, a half-life limit of more than $10^{26}$ y will be 
reached, corresponding to a mass of $\sim$ 0.060 eV. In the second step, 370 crystals (for a total mass of $\sim$ 1 ton) 
will be placed within the Borexino apparatus. In this case, the sensitivity will be greater than $10^{27}$ y, and a mass
limit in the range of $\sim$ 0.020 eV should be reached.   
\end{itemize}
\item
{\bf{Tellurium}} \\
The CUORE (Cryogenic Underground Observatory for Rare Events) project consists of a series of 
experiments with massive cryogenic detectors to investigate rare processes, in particular the DBD reaction. 
The final setup of the detector is still under analysis. A possible way-out is a structure with 988 cubic natural TeO$_{2}$ crystals (5 cm size and a mass of 760 g each), 
arranged in 19 columns with 12 flours of 4 crystals, operating at T = 0.01 K. The total mass will be of 750 kg of
TeO$_2$, which corresponds to 203 kg of $^{130}$Te. Crystals will be separated by a few mm of material.
The expected background is $\sim$ 
0.001 counts/(keV kg y), with an energy resolution of $\sim$ 5 keV at the Q value (2.529 MeV).
The main background component is due to a surface contamination. A great advantage of this experiment is the high 
natural abundance of $^{130}$Te, moreover cosmogenic activities within the crystals are reduced. The cryostat 
is also shielded by Roman lead having an activity lower than 4 mBq/kg, surrounded by modern lead, whose activity is  $\sim$
16 Bq/kg.
The estimated sensitivity is $\sim$ $\sqrt{t}\,\cdot 10^{26}$ y, where t is the measurement time in year. 
After 1 year of data taking, the mass limit of 0.04 - 0.15 eV will be available, \cite{Arnaboldi2003, Fiorini2004}.

\item
{\bf{Xenon}} \\
Two main projects plan to look for $\rm ^{136}Xe$.
\begin{itemize}
\item The EXO (Enriched Xenon Observatory) experiment should use a new approach that combines quantum 
optics techniques with radiation detectors, aiming to detect single Ba$^{+}$ ions, via resonant excitation with a set of 
lasers, in the final state of $^{136}$Xe DBD. It will use several tons of $^{136}$Xe, enriched up to 80\%. The energy
resolution will be  $\sim$ 2\% at 2.5 MeV. 
Two different techniques are still under analysis: high pressure gas Time Projection Chamber and liquid Xenon scintillator, which
offers much more compact sizes. The EXO collaboration is still preparing a 200 kg prototype detector, \cite{Danilov2000}. 
Recently, the required energy resolution has been obtained by simultaneous measurements of ionisation and scintillation 
light.  
With a 1 ton detector and 5 y of measurements, a sensitivity of 8$\cdot 10^{26}$ y, or in the mass interval
0.050 - 0.140 eV, is expected. The detector should be installed at WIPP Laboratories in Carlsbad. 
\item The XMASS (Xenon neutrino MASS detector) experiment will take place at Kamioka Underground Laboratory, Japan.
The detector will use liquid Xenon viewed by photomultipliers. In fact, liquid Xenon is a good scintillator, and has a
high Z value, density and boiling point. Moreover, Xenon allows purification to take place during the operations. 
An intense R\&D phase with a 100 kg detector has confirmed the reliability of 
vertex and energy reconstruction, the self shielding power for $\gamma$ rays. Then, it has allowed to measure environmental
background and internal radioactive impurity. Therefore, a 800 kg detector is currently under 
construction. The final step will be a 10 ton
detector, which should reach a sensitivity of $\sim 8\cdot10^{21}$ y for $2\nu$-DBD and 
$\sim 3.3\cdot10^{26}$ y for $0\nu$-DBD, which implies a neutrino mass limit of 0.03 - 0.09 eV without 
enriched materials, \cite{Minamino2004}.
\end{itemize}

\item
{\bf{Neodymium}} \\
The DCBA (Drift Chamber Beta-ray Analyser) experiment is searching for $0\nu$-DBD reaction from $^{150}$Nd. 
The parameters of this tracking detector, which should be composed by 20 kg of enriched $^{150}$Nd, are under analysis
at KEK in Japan,  \cite{Ishihara2003}. After a test apparatus for technical development, a standard module, which will 
use natural Nd as source, will check the feasibility of the whole apparatus, which will consist of a 100 module array with enriched Nd.

\end{enumerate}

Table \ref{table future projects} summarizes the expected sensitivities.

\begin{table*}
\centering
\caption[]
{Expected sensitivities and effective neutrino mass limits for future projects. 
For $\langle m_{\nu} \rangle$ calculations, the nuclear matrix elements from \cite{Staudt1990} has been used.}
\label{table future projects}
\begin{tabular}{llllll}  \hline
Experiment   &Isotope   &Mass   & T$_{1/2}^{0\nu}$  &$\langle m_{\nu} \rangle$ & References\\
             &          &(kg)   & $10^{26}$ y    & (eV)  &  \\
\hline
CAMEO 1+CTF  &$^{116}$Cd &$10^3$      &10  & 0.060        & \cite{Bellini2001} \\
CAMEO 2+Borexino&$^{116}$Cd &$10^3$  &100  & 0.020        & \cite{Bellini2001} \\
CANDLES     &$^{48}$Ca &$> 10^3$ &$>1$      & 0.030      & \cite{Umehara2003} \\ 
COBRA      &$^{130}$Te &10 &0.01           & 0.240     & \cite{Zuber2001} \\
CUORE      &$^{130}$Te &750 &7             & 0.027      & \cite{Arnaboldi2002} \\
DCBA       &$^{150}$Nd &20  &0.15          & 0.035      & \cite{Ishihara2003} \\ 
EXO        &$^{136}$Xe &$10^3$ &8          & 0.052      & \cite{Danilov2000} \\
GEM        &$^{76}$Ge  &$10^3$ &70         & 0.018      & \cite{Zdesenko2001} \\
GENIUS     &$^{76}$Ge  &$10^3$ &100        & 0.015      & \cite{Klapdor2001c} \\
MAJORANA   &$^{76}$Ge  &500 &40            & 0.030      & \cite{Aalseth2004} \\
MOON       &$^{100}$Mo  & few $10^3$&10    & 0.036      & \cite{Ejiri2000} \\
XMASS      &$^{136}$Xe  &$10^4$ &3         & 0.086      & \cite{Moriyama2001} \\
\hline
\end{tabular}
\end{table*}

\subsection{Other proposals}
In the past few years, other approaches have been proposed and developed:
\begin{itemize}
\item The use of CTF and/or Borexino apparatuses as $0\nu$-DBD detector, see \cite{Bellini2001, Caccianiga2001}
\item The use of the SNO detector, after the end of solar neutrinos experiments, filled by a 1\% loaded liquid scintillator.
An extension of the present underground laboratory (SNO-LAB) has been also proposed to the physics community.
\item The systematic studies of $0\nu$-DBD reaction toward excited states of daughter nuclei, see for instance
\cite{Suhonen2000, Simkovic and Faessler 2002, Barabash2003, Barabash2004a}, and $via$ $\beta^+$ decays, even if this process offers a reduced phase space 
and long half-life, see \cite{Danevich2004}. Another opportunity is given by the radiative neutrinoless double electron capture, see \cite{Sujkowski2004}, which could be an intriguing sector for isotopes like $^{112}$Sn. The use of doped neodymium crystals has been also recently analysed, see \cite{Danevich2004a}. Unfortunately, the present 
knowledge of the properties of such nuclei is rough. A strong improvement in nuclear matrix element calculations for both the cases is needed, see \cite{Kiel2003}.
\item Search for $0\nu$-DBD reactions of initially unstable nuclei, see \cite{Tretyak2004}.
\end{itemize}   
  
\section{Discussion and perspectives}

At present, the most powerful results and limits in $0\nu$-DBD experiments have been obtained by detectors
using $^{76}$Ge as source (IGEX and HM). They have reached an upper half-life limit of $\sim$ 1.6$\cdot 10^{25}$y, 
which corresponds to a sensitivity of $\sim$ 0.3 eV, even if nuclear matrix element calculations induce a 
significant uncertainties in the mass value. Moreover, in spring 2004 the experimental discovery of $0\nu$-DBD reaction in
$^{76}$Ge has been claimed, with a half-time  $\sim$ 1.2$\cdot 10^{25}$ y. All these values are beyond the present 
capability of other experiments, which cannot confirm or deny these results.

The deduced mass value is close to cosmological limits on neutrino masses: WMAP collaboration produces an upper 
limit of 0.23 eV per neutrino flavour, but this result is strongly dependent on the hypotheses assumed in the calculation, and a more conservative limit is $\sim$ 0.60 eV. 

Other currently running $0\nu$-DBD experiments, like CUORICINO and NEMO-3, have the possibility to reach 
and verify this limit within few years of data taking. The KATRIN tritium beta decay apparatus should check 
this energy region and confirms the recently claimed $0\nu$-DBD experimental discovery. 

\subsection{Has the $0\nu$-DBD reaction been discovered?}

In 2001, some members of the HM collaboration claimed evidence of $0\nu$-DBD at a level of 
2.2-3.1 $\sigma$, by assuming a flat background in a small region centred around the Q peak, \cite{Klapdor2001}. 
Other similar peaks are present in the selected region, and a refined analysis over wide energy region with 
the inclusion of $^{214}$Bi lines reduced the previous effect at no more than 1.5 $\sigma$. 
It has to be stressed that the remaining members of HM collaboration did not claim the $0\nu$-DBD  
discovery in that dataset.

At the beginning of 2004, Klapdor-Kleingrothaus and coll. strongly confirmed the evidence of 
$0\nu$-DBD process after an analysis on the whole dataset between August 1990 and May 2003, \cite{Klapdor2004}. 
The main characteristics of the experiment and their results are the following:
\begin{itemize}
\item Active volume of 10.96 kg of HP p-type $^{76}$Ge, enriched at 86-88 \% level;
\item Whole duty cycle of $\sim$ 80\%;
\item Collected statistics of 71.7 kg $\cdot$ y;
\item Energy resolution at a level of 0.2\%;
\item Background of 0.113 $\pm$ 0.007 counts/(keV kg y) in the $0\nu$-DBD region, 
around the Q peak, which occurs at 2039.006 $\pm$ 0.050 keV;
\item About $10^6$ events registered since 1995, with new and improved setup;
\item A signal of 28.75 $\pm$ 6.86 events;
\item When the nuclear matrix calculations given in \cite{Staudt1990} are used, the 3$\sigma$ range results are within the
interval (0.69 - 4.18)$\cdot 10^{25}$ y for the half-life, with an effective neutrino mass of 0.24 - 0.58 eV, 
at 4.2 $\sigma$. The best fit values are 1.19$^{+0.37}_{-0.23}\cdot 10^{25}$ y, and 0.44 eV, respectively. 
\end{itemize} 

This result is unique and under the analyses of the DBD community.  
In any case, the identification of a genuine signal in an energy region where background counts are at 
a similar level is a very difficult task. A confirmation by other experimental groups is needed, even if 
just the combination of IGEX and HM $^{76}$Ge crystals should verify such results within 2-3 years, by using the same 
decaying nucleus.  

Another consequence of the claimed detection of $0\nu$-DBD reaction is the necessity to right evaluate the systematic 
uncertainties, which have usually been estimated as a negligible contribution to the genuine signal.

If the experiments will confirm the quoted half-life and mass values, all particle physics should be renewed, see for instance the references in \cite{Klapdor2004} for an overview.

\section{Conclusions}

Recent experimental results have shown the neutrinos are changing their flavours when they travel between 
sources and detectors. The range of values for square mass differences and mixing angles have been deduced.
Therefore, neutrino mass eigenstates (at least, two or three, depending on the number of neutrino families) do have a non-zero mass.\\
Unfortunately, no further news concerning the mass of each term and the neutrino 
Dirac or Majorana nature is allowed. Neutrinoless DBD is a unique process which could offer an answer 
to these questions.

New results from cosmology and neutrino mass beta decay based experiments should check the degenerate solution, whereas 
long baseline experiments could confirm inverted hierarchy spectrum. The KATRIN experiment can test the 0.25 eV mass region 
within few years.

We also remind that if an established neutrino mass limit is searched for, very different half-life values have to be measured, depending on the selected isotope. 
As an example, if $\left|\langle m^{\nu}\rangle\right| <$ 0.04 eV, the half-life values vary from few $10^{25}$y for $^{150}$Nd up to some $10^{27}$y for $^{48}$Ca, $^{76}$Ge, $^{116}$Cd and $^{136}$Xe, but the selection of a candidate isotope depends on many other parameters.

The actual stronger constraints on neutrino masses are from the cosmological sector, based on WMAP, 2dFGRS, and Lyman-$\alpha$ analyses. The obtained limit $m_1 + m_2 + m_3 <$ 0.70 - 1.70 eV implies $\left|\langle m^{\nu}_{ee} \rangle\right| <$ 0.23 - 0.60 eV, which has to be compared with the limit quoted in \cite{Vogel2002}.

A common task for all $0\nu$-DBD experimental groups is a further suppression of background events, such as environmental radioactivity, cosmic component, internal contamination. Only a significant reduction of such reactions and an enhancement of genuine signals will allow to measure longer half-lifes, and consequently, smaller neutrinos mass intervals.

The next generation of $0\nu$-DBD detectors, which have an expected sensitivity down to 0.01 eV, should allow the identification of the Dirac or Majorana nature of the neutrino, for the cases of the degenerate and inverted mass spectra, see \cite{Bilenky2004}:
\begin{itemize}
\item If the $0\nu$-DBD reaction will be not detected by next-generation experiments and the effective neutrino mass is lower than 0.045 eV, then a normal neutrino mass hierarchy occurs. Consequently, massive neutrinos can be either Dirac or Majorana particles.
\item If the $0\nu$-DBD reaction will be observed and the effective mass is greater than 0.045 eV, then the normal hierarchy is excluded.  
\item If the $0\nu$-DBD reaction will be detected, and $0.4\,\sqrt{\Delta\, m_{\mathrm {atm}}^{2}} 
\leq \left|\langle m_{\nu}\rangle\right|\leq \sqrt{\Delta\, m_{\mathrm {atm}}^{2}}$, then an inverted mass hierarchy occurs.
\item If the $0\nu$-DBD reaction will be measured, and  
$\left|\langle m_{\nu}\rangle\right|\gg \sqrt{\Delta\, m_{\mathrm {atm}}^{2}}$, then the mass spectrum is almost degenerate.
\item If future beta decay experiments or cosmological measurements will offer new limits, then the effective neutrino mass will be deduced from the relation 
$0.4\,m_1 \leq \left|\langle m_{\nu}\rangle\right|\leq m_1$.
If the $0\nu$-DBD process will not be observed, or if the effective Majorana neutrino mass is out of this range, neutrinos are Dirac particles or other mechanisms producing total lepton number violation are required. 
\end{itemize}

\section{Acknowledgements}

Many thanks to V. Antonelli (INFN, Milan) and A.S. Barabash for the helpful comments.


\end{document}